\documentclass[12pt]{article}

\usepackage{a4wide}
\usepackage[dvips]{graphicx}
\graphicspath{{/home/kroll/Tex/SDME/Figs-sdme/}}

\newcommand{\nn}{\nonumber}
\newcommand{\be}{\begin{equation}}
\newcommand{\ee}{\end{equation}}
\newcommand{\ba}{\begin{eqnarray}}
\newcommand{\ea}{\end{eqnarray}}
\newcommand{\ci}[1]{\cite{#1}}
\def\vk{{\bf k}_{\perp}}

\def\vbs{{\bf b}}
\def\vb0{{\bf b}_0}

\def\als{\alpha_s}

\def\mev{\,{\rm MeV}}
\def\gev{\,{\rm GeV}}

\def\xbj{x_{\rm Bj}}
\newcommand{\sla}{\hspace*{-0.20cm}/}
\newcommand{\Sla}{\hspace*{-0.16cm}/}

\newcommand{\wf}{wavefunction}
\newcommand{\lsim}{\raisebox{-4pt}{$\,\stackrel{\textstyle
                                                         <}{\sim}\,$}}
\newcommand{\gsim}{\raisebox{-4pt}{$\,\stackrel{\textstyle
                                                         >}{\sim}\,$}}

\newcommand{\tw}{\textwidth}
                                                         
\newcommand{\req}[1]{(\ref{#1})}
\def\xb{\bar{x}}

\def\={\,=\,}
\def\eps{\epsilon}
\def\veps{\varepsilon}

\begin{document}
\thispagestyle{empty}
\begin{flushright}
WU B 07-07 \\
September, 16 2007\\[20mm]
\end{flushright}

\begin{center}
{\Large\bf The role of the quark and gluon GPDs in hard vector-meson
  electroproduction} \\
\vskip 15mm

S.V.\ Goloskokov
\footnote{Email:  goloskkv@theor.jinr.ru}
\\[1em]
{\small {\it Bogoliubov Laboratory of Theoretical Physics, Joint Institute
for Nuclear Research,\\ Dubna 141980, Moscow region, Russia}}\\
\vskip 5mm

P.\ Kroll \footnote{Email:  kroll@physik.uni-wuppertal.de}
\\[1em]
{\small {\it Fachbereich Physik, Universit\"at Wuppertal, D-42097 Wuppertal,
Germany}}\\

\end{center}

\vskip5mm
\begin{abstract}
\noindent Electroproduction of light vector mesons is analyzed on the basis of
handbag factorization. The required generalized parton distributions 
are constructed from the CTEQ6 parton distributions with the help of 
double distributions. The partonic subprocesses are calculated within 
the modified perturbative approach. The present work extends our 
previous analysis of the longitudinal cross section to the transverse 
one and other observables related to both the corresponding
amplitudes. Our results are compared to recent experimental findings 
in detail.
\end{abstract}


\section{Introduction}
\label{sec:introduction}
\begin{figure}[t]
\begin{center}
\includegraphics[width=.35\textwidth,bb=113 391 312 530,%
clip=true]{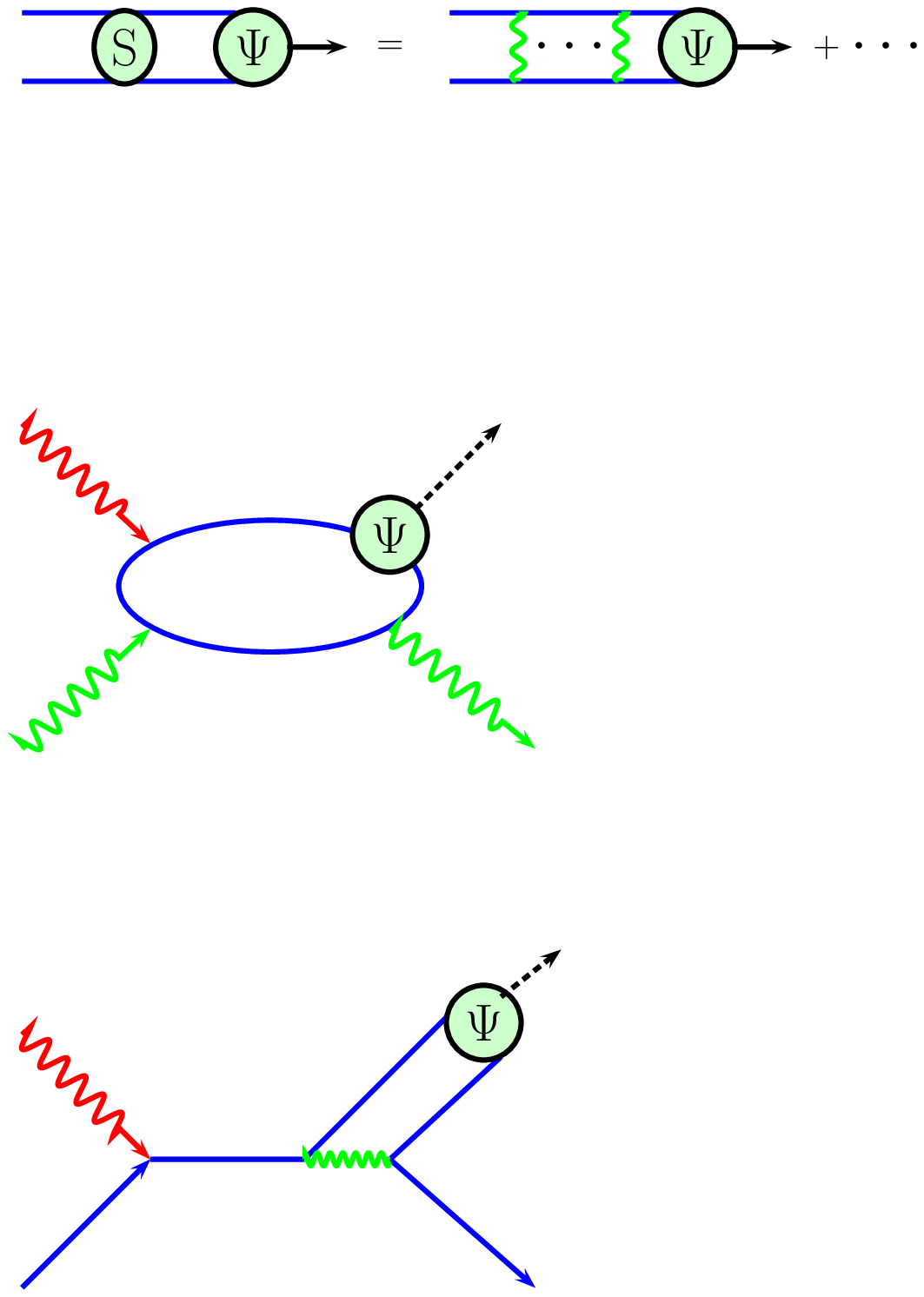}\hspace*{0.6cm}
\includegraphics[width=.35\tw,bb=113 221 299 340,%
clip=true]{meson-graph.ps} 
\end{center}
\caption{Typical lowest order Feynman graphs for the $\gamma^* g\to
  Vg$ (left) and $\gamma^* q\to Vq$ (right) subprocesses
  of meson electroproduction.} 
\label{fig:feynman}
\end{figure}

In a previous work \ci{first} we analyzed electroproduction of light
vector mesons ($V=\rho^0, \phi, \omega$) at HERA kinematics within the 
handbag factorization scheme which is based on generalized parton
distributions (GPDs) and hard partonic subprocesses. The latter  
are calculated within the modified perturbative approach \ci{botts89}
in which the quark transverse momenta are retained. The emission and 
reabsorption of partons from the proton is still treated in collinear 
approximation. In the kinematical region accessible to the HERA 
experiments that is characterized by  Bjorken-$x$ ($\xbj$) of the 
order $10^{-3}$, it is not unjustified to restrict oneself to the
gluonic subprocess $\gamma^* g\to Vg$ and the associated gluonic GPD 
$H^g$. In a recent paper \ci{second} we extended that analysis to  
larger values of $\xbj$ ($\lsim 0.2$) such as accessible to the HERMES 
experiment, but restricting ourselves to the analysis of the least 
model-dependent amplitude, namely the one for transitions from 
longitudinal polarized virtual photons to vector mesons polarized in 
the same manner, $\gamma^*_{\,L} p\to V_{\,L}p$.  This analysis 
necessitates the inclusion of the quark subprocesses 
$\gamma^* q \to Vq$ (see Fig.\ \ref{fig:feynman}) and the associated 
GPDs for sea and valence quarks.   

Here, in this work we are going to complete the analysis of
vector-meson electroproduction by studying the amplitude for 
transversely polarized photons and mesons, $\gamma^*_{\,T}p\to
V_{\,T}p$. The basic idea of modeling this amplitude has been already
described in Ref. \ci{first} for the gluonic contribution. The
extension to the quark contribution is straightforward. The crucial 
point at issue with the transverse amplitude is that the quark 
transverse momenta which are retained in the modified perturbative 
approach in order to suppress configurations with large transverse 
separations of the quark and antiquark forming the meson, regularize 
the infrared singularities occurring in the transverse subprocess 
amplitude in collinear approximation \ci{man,annikin} at the same time. 
These special configurations seem to be responsible for the excess 
of the leading-twist contribution to the longitudinal cross section 
over experiment. Indeed, taking into account the quark transverse 
momenta fair agreement between theory and experiment is achieved 
\ci{second}. We are going to apply this modified handbag approach to 
vector-meson electroproduction for energies, $W$, in the photon-proton 
center of mass system (c.m.s.) between about 5 and $170\,\gev$ and 
photon virtualities, $Q^2$, between about 3 and $100\,\gev^2$ while 
$\xbj$ is less than $\lsim 0.2$. We will compare our results in detail 
with recent data from HERA, COMPASS and HERMES. 

The GPDs $H$ for quarks and gluons which dominate the process of
interest for unpolarized protons at small $\xbj$, are constructed from
the CTEQ6 parton distribution functions (PDFs) \ci{cteq6}   
through double distributions \ci{mul94, rad98}. Applying the same model
to the GPDs $\widetilde{H}$ we are also going to estimate the so-called
unnatural parity amplitudes and to study their implications on spin
density matrix elements (SDME) and the double spin asymmetry $A_{LL}$
describing the correlation of the helicities of the beam and target
particles.
 
In Ref.\ \ci{DFJK4} the electromagnetic form factors of the nucleon
have been used to extract the zero-skewness GPDs $H$, $\widetilde{H}$
and $E$ for valence quarks. The forward limit of $E^{\,a}_{\rm val}$, 
the analog of the PDFs, determined in Ref.\ \ci{DFJK4}, can be utilized 
for the construction of $E^{\,a}_{\rm val}$ at non-zero skewness with
the double distribution model. With these GPDs at disposal we 
will also estimate the SDMEs measurable with a transversely polarized 
target as well as the transverse spin asymmetry $A_{UT}$.

The plan of the paper is as follows: In Sect.\ \ref{sec:handbag} we
will sketch the modified handbag approach. In Sect.\ \ref{sec:GPD} the
double distribution model for the GPD $H$ is described. The results on 
cross sections and SDMEs obtained within the handbag approach are presented
in Sect.\ \ref{sec:trans} and \ref{sec:sdme}, respectively. Sect.\
\ref{sec:Htilde} is devoted to an estimate of the role of the GPDs
$\widetilde{H}$ and Sect.\ \ref{sec:hel-flip} to that of the GPDs
$E$. We will conclude this paper by a summary (Sect.\ \ref{sec:summary}).

\section{The handbag approach}
\label{sec:handbag} 
We are interested in vector-meson electroproduction in a kinematical
region characterized by large $Q^2$ and large $W$ but small $\xbj
(\lsim 0.2)$ and small invariant momentum transfer $-t$. In the
handbag approach, the amplitudes for the process $\gamma^*p\to Vp$
which can be extracted from vector-meson electroproduction applying
the one-photon approximation, factorize into partonic
subprocesses (see Fig.\ \ref{fig:feynman}) and GPDs comprising the
soft, non-perturbative QCD. At large $Q^2$ the amplitude for 
$\gamma_{\,L}^*p\to V_{\,L}p$ dominates and factorization has been 
shown to hold for it rigorously \ci{rad96,col96}. The amplitudes for 
other photon-meson transitions are suppressed by inverse powers of
$Q$. Besides the longitudinal amplitude we will consider only the 
transverse one, $\gamma_{\,T}^*p\to V_{\,T}p$ in this work which is
the most important one of the suppressed amplitudes at small $-t$.
Proton helicity flip is suppressed by $\sqrt{-t}/2m$ 
($m$ being the proton's mass) and can be neglected in calculations of 
cross sections and SDMEs obtained with unpolarized protons. In Sect.\ 
\ref{sec:hel-flip} we will however estimate the size of the proton 
helicity-flip amplitudes explicitly. It will turn out that these 
amplitudes are indeed small.

In the region of small $\xbj$ the dominant contributions are provided
by the GPD $H$. To the proton helicity non-flip amplitude the GPDs
contribute in the combination
\be
        H - \frac{\xi^2}{1-\xi^2}\, E
\label{eq:H-E}
\ee
where the skewness $\xi$ is related to $\xbj$ by
\be
\xi \simeq \frac{\xbj}{2-\xbj}\,\big[1+m_V^2/Q^2\,\big]\,.
\label{xi-xbj}
\ee
Here, $m_V$ denotes the mass of the vector meson. The GPD $E$ can
therefore safely be ignored since it is not expected that it is much 
larger than the GPD $H$, see Sect.\ \ref{sec:hel-flip} where we will
take up this issue again. The GPD $\widetilde{H}$  only contributes to
the transverse amplitude and can also be neglected in calculations of
the cross sections. In Sect.\ \ref{sec:Htilde} we will return to this 
problem and estimate the size of its contribution. Since there is no 
parton helicity flip in the partonic subprocesses to the accuracy we 
are calculating them, the parton helicity-flip GPDs \ci{diehl03} do
not occur.
  
The contributions from $H$ to the $\gamma^*p\to Vp$ amplitudes read
($i=g,q$, $x_g=0$, $x_q=-1$)
\be
{\cal M}_{\mu +, \mu +}^{Ni}(V) \= \frac{e}{2} \sum_a 
         e_a{\cal C}_V^{\,a}\, \int_{\xb_i}^1\, d\xb \, 
 \sum_{\lambda} {\cal H}^{Vi}_{\mu\lambda,\mu\lambda}(\xb,\xi,Q^2,t=0)
             \,H^i(\xb,\xi,t)\,.
\label{H-amp}
\ee
The first sum runs over the quark flavors $a$ and $e_a$ denotes the quark
charges in units of the positron charge $e$. The non-zero flavor weight 
factors, ${\cal C}_V^a$, read 
\be
{\cal C}_\rho^{\,u}\=-{\cal C}_\rho^{\,d} \= {\cal C}_\omega^{\,u} 
\={\cal C}_\omega^{\,d}\=1/\sqrt{2}\,, \qquad {\cal C}_\phi^{\,s}\= 1\,.
\label{flavor}
\ee
The explicit helicities in \req{H-amp} refer to the proton while $\mu$
is the helicity of the photon and meson and $\lambda$ that of the partons
participating in the subprocess. Only the $t$ dependence of the GPDs
is taken into account in the amplitudes \req{H-amp}. That of the
subprocess amplitudes ${\cal H}$ provides corrections of order $t/Q^2$
which we neglect throughout this paper. In contrast to the subprocess
amplitudes the $t$ dependence of the GPDs is scaled by a soft parameter, 
actually by the slope of the diffraction peak. 

There is a minimal value of $-t$ allowed in the process of interest
\be
t_{\rm min}\= -4m^2\,\frac{\xi^2}{1-\xi^2}\,.
\ee
As other effects of order $\xi^2$ (e.g.\ the GPD $E$) the $t_{\rm
  min}$ effect is neglected. We note in passing that our helicities
are light-cone ones which naturally occur in the handbag approach. The
differences to the usual c.m.s. helicities are of order
$m\sqrt{-t}/W^2$ \ci{diehl03} and can be ignored in the kinematical
region of interest in this work. 

The full amplitude is given by a superposition of the gluon and quark
contributions
\be
{\cal M}^N \= {\cal M}^{Ng} + {\cal M}^{Nq}
\label{amp}
\ee
and it is normalized such that the partial cross sections for $\gamma^*p\to Vp$
read ($\Lambda$ is the usual Mandelstam function)
\be
\frac{d\sigma_{L(T)}}{dt} \= \frac1{16\pi (W^2-m^2) 
         \sqrt{\Lambda(W^2,-Q^2,m^2)}}\,|{\cal M}^N_{0(+)+,0(+)+}|^2\,,
\label{sigma}
\ee
which holds with regard to the above-mentioned simplifications. The
cross sections integrated over $t$ are denoted by $\sigma_L$ and
$\sigma_T$. The full (unseparated) cross section for $\gamma^*p\to Vp$ is
\be
\sigma \= \sigma_T + \veps \sigma_L\,,
\ee
in which $\varepsilon$ is the ratio of longitudinal to transverse
photon fluxes. The power corrections of 
kinematical origin given in Eq.\ \req{xi-xbj} and in the phase space 
factor \req{sigma} are taken into account by us. With the exception of 
these kinematical effects hadron masses are omitted otherwise.
In the expression \req{sigma} for the cross section the
symmetry relation
\be
{\cal M}^{Ni}_{-\mu\nu^\prime,-\mu\nu} \= 
             {\cal M}^{Ni}_{\mu\nu^\prime,\mu\nu}
\label{sym-N}
\ee
has been used which is an obvious consequence of the definition
\req{H-amp} and parity conservation. This symmetry relation coincides 
with the one that holds for natural parity  exchanges; we therefore
mark this amplitude by a superscript $N$. Since the contributions from
$\widetilde{H}$ to the amplitudes obey the relation
\be
{\cal M}^{Ui}_{-\mu\nu^\prime,-\mu\nu} \= -
             {\cal M}^{Ui}_{\mu\nu^\prime,\mu\nu}\,,
\label{sym-U}
\ee 
there are no interference terms between ${\cal M}^{Ni}$ and ${\cal M}^{Ui}$ 
in the observables for unpolarized electroproduction of vector mesons as can
easily be shown. The relation \req{sym-U} is also obtained  for the 
exchange of a particle with unnatural parity. In analogy to the 
contributions from $H$ the amplitudes related to $\widetilde{H}$
are marked by a superscript $U$. The contribution $\mid {\cal M}^{U}\mid^2$ 
is neglected in \req{sigma}. 

Let us now turn to the discussion of the subprocess amplitudes. As is
well-known, for the kinematics accessible to current experiments, the
handbag amplitude evaluated in collinear approximation, overestimates   
the longitudinal cross section although with the tendency of
approaching experiment with increasing $Q^2$ \ci{first,kugler,kugler-err}.
One may wonder whether higher order perturbative QCD corrections to
the subprocesses may cure that deficiency. However, this does not seem
to be the case. NLO corrections \ci{ivanov04,kugler07} are very large
due to BFKL-type logarithms $\sim \ln{1/\xi}$ and cancel to a large
extent the LO term at low $Q^2$ and low $\xbj$. A recent attempt 
\ci{ivanov07}  to resum higher orders with methods known from deep 
inelastic lepton-nucleon scattering \ci{catani} seems to indicate that
the sum of all higher order corrections to the LO term is not large. 
In view of this unsettled situation it seems to be reasonable to
proceed along the lines advocated in Refs.\ \ci{first,second} by 
simply using the LO result and add power corrections, especially since 
such corrections are anyway needed in order to account for the large 
transverse cross section $\sigma_{\,T}$. Once the higher order 
perturbative corrections are better understood within the modified
perturbative approach one may add them to the LO results. As long as 
they are of reasonable magnitude there is no principal difficulty in
this. It may merely be necessary to readjust the GPDs and the meson 
wave functions appropriately. 

As in our papers \ci{first,second} we will model the required power
corrections by employing the modified perturbative approach
\ci{botts89} in the calculation of the subprocesses. In this approach
the transverse momenta of the quark and antiquark, $\vk$ (defined with 
respect to the meson's momentum), entering the meson are kept. In 
contrast to the situation at the mesonic vertex, the partons entering 
the subprocess, are viewed as being emitted and reabsorbed by the
proton collinearly. This scenario is supported by the fact that
the GPDs describe the full proton and their $\vk$ dependence therefore
reflects the proton's charge radius 
($\langle \vk^2\rangle^{1/2}\simeq 200\,\mev$) while the meson is 
generated through its compact valence Fock state with a r.m.s. $\vk$ 
of about $500\,\mev$ \ci{jakob,jakob-err}. Instead of a meson's distribution 
amplitude allowance is to be made for a meson light-cone \wf{} 
$\Psi_{VL}(\tau,k_\perp)$ in the modified perturbative approach 
\ci{jakob,jakob-err}. Here $\tau$ is the fraction of the light-cone plus 
component of the meson's momentum, $q'$, the quark carries; the 
antiquark carries the fraction $\bar{\tau}\=1-\tau$. Quark transverse 
momenta are accompanied by gluon radiation. In Ref.\ \ci{botts89} the 
gluon radiation has been calculated in the form of a Sudakov factor 
${\rm exp}[-S(\tau,{\bf b},Q^2)]$ to next-to-leading-log approximation using 
resummation techniques and having recourse to the renormalization
group. The quark-antiquark separation, $\vbs$, in configuration space 
acts as an infrared cut-off parameter. Radiative gluons with wave
lengths between the infrared cut-off and a lower limit (related to
the hard scale $Q^2$) yield suppression, softer gluons are part of the 
meson \wf{} while harder ones are an explicit part of the subprocess 
amplitude. Congruously, the factorization scale is given by the 
quark-antiquark separation, $\mu_F=1/b$, in the modified perturbative 
approach. For more details we refer to Refs.\ \ci{botts89,second}.  

Since the resummation of the logarithms involved in the Sudakov factor can
only efficiently be performed in the impact parameter space \ci{botts89}
we have to Fourier transform the lowest-order subprocess amplitudes
to that space and to multiply them with the Sudakov factor there. This
leads to 
\be
{\cal H}^{Vi}_{\mu\lambda,\mu\lambda} \= \int d\tau d^2b\, 
         \hat{\Psi}_{VL}(\tau,-\vbs)\, 
      \hat{\cal F}^{\,i}_{\mu\lambda,\mu\lambda}(\xb,\xi,\tau, Q^2,\vbs)\, 
         \als(\mu_R)\,{\rm exp}{[-S(\tau,\vbs,Q^2)]}\,.
\label{mod-amp}
\ee
The two-dimensional Fourier transformation between the canonically conjugated
$\vbs$ and $\vk$ spaces is defined by 
\be
    \hat{f}(\vbs) \= \frac{1}{(2\pi)^2} \int d^{\,2}\, \vk\, \exp{[-i
                                  \,\vk\cdot\vbs\,]}\; f(\vk)\,.
\label{fourier}
\ee
The renormalization scale $\mu_R$ is taken to be 
the largest mass scale appearing in the hard scattering amplitude,
i.e. $\mu_R=\max\left(\tau Q, \bar{\tau} Q,1/b\right)$. Since the 
bulk of the handbag contribution to the amplitudes is accumulated in
regions where $\mu_R$ is smaller than $3\,\gev$ we have to deal with
three active flavors. A value of $220\,\mev$ for 
$\Lambda_{QCD}$ is used in the Sudakov factor and in the evaluation 
of $\als$ from the one-loop expression. 

The hard scattering kernels ${\cal F}^{i}$ or their Fourier
transform $\hat{{\cal F}}^{i}$ occurring in Eq.\ \req{mod-amp}, are
computed from the pertinent Feynman graphs, see Fig.\ \ref{fig:feynman}. 
The result for the gluonic subprocess is discussed in some detail in
Ref.\ \ci{first} and we refrain from repeating the lengthy expressions
here. For quarks, on the other hand, the hard scattering kernel for
longitudinally polarized photons and mesons reads 
\be
{\cal F}^{\,q}_{0+,0+}+{\cal F}^{\,q}_{0-,0-} \= - C_F \sqrt{\frac{2}{N_c}}\,
    \frac{Q}{\xi}\,\left[T_s-T_u\right]\,,
\label{kernel-long}
\ee
where $N_c$ denotes the number of colors and $C_F=(N_c^2-1)/(2N_c)$ is
the usual color factor. For convenience we only quote the sum over
the quark helicities since this is what appears in \req{H-amp}. The  
denominators of the parton propagators read
\be
T_s\=\frac1{k_\perp^2 -\tau(\xb-\xi)Q^2/(2\xi) - i\eps}\,, \qquad
T_u\=\frac1{k_\perp^2 + \bar{\tau}(\xb+\xi)Q^2/(2\xi)-i\eps} \,.
\label{q-kernel}
\ee
In both the quark and the gluon propagators we only retain $k_\perp$
in the denominators of the parton propagators where it plays a crucial
role. Its square competes with terms $\propto \tau (\bar{\tau}) Q^2$ 
which become small in the end-point regions where either $\tau$ or
$\bar{\tau}$ tends to zero.

While for longitudinally polarized vector mesons the quark transverse
momentum is only needed for the suppression of the leading-twist 
contribution (the $\vk\to 0$ limit), it plays an even more important 
role in the case of transverse polarization. In the collinear limit
the spin \wf{} of the meson is $\Gamma_V \propto q\sla^\prime 
\eps\sla_V(\pm 1)$ where $\eps_V$ denotes the polarization vector of 
the meson. It however leads to a vanishing contribution to the
subprocess amplitude since the number of $\gamma$ matrices in the
Dirac trace~\footnote{
For the quark subprocess the trace includes the hadronic matrix
element that defines the GPDs and is $\propto \gamma^+$ or $\propto
\gamma^+\gamma_5$.}
that gives the hard scattering kernel, is odd (see Fig.\ \ref{fig:feynman}
)~\footnote{
Note that for longitudinally polarized vector mesons $\eps(0) \simeq
q^\prime/m_V$. Hence, the spin \wf{} is $\propto q\Sla^\prime$ in this
case up to mass corrections. A mass term $\propto
m_V\eps\hspace*{0.003\tw}\Sla\hspace*{-0.003\tw}_V$ in
the spin \wf{} for transversely polarized vector mesons has been
investigated \ci{man}. In collinear approximation this term leads to
an infrared singular twist-3 contribution. Such mass terms are
neglected by us.}.
If one allows for quark transverse momenta a second term in the
covariant spin \wf{} appears
\be
\Delta \Gamma_V^\nu K_\nu \= \frac1{\sqrt{2}M_V}\,
\{q\sla'\,\eps\sla_V,\gamma_\nu\}\,K^\nu\,,
\label{spin-wf}
\ee
for which the number of $\gamma$ matrices in the Dirac trace is even. 
In eq.\ \req{spin-wf} $M_V$ is a soft parameter of the order of the 
vector-meson mass, e.g.\ a typical constituent quark mass. The 
transverse momentum four-vector $K=[0,0,\vk]$ is suitably defined as 
the quark-antiquark relative momentum and represents one unit of 
orbital angular momentum in a covariant manner \ci{first}. Expanding 
now the hard scattering kernel for transversely polarized vector
mesons, one finds
\ba
{\cal F}^{\,i}_{+\lambda,+\lambda}(\xi, \xb, \tau, Q^2, K) &=&
          {\cal F}^{\,i}_{+\lambda,+\lambda} (\xi, \xb, \tau, Q^2, \vk^2)
              + \Delta {\cal F}^{\,i\, \nu}_{+\lambda,+\lambda}
                     (\xi, \xb, \tau, Q^2, \vk^2) K_\nu\nn\\ 
&+& \Delta {\cal F}^{\,i\,\nu\mu}_{+\lambda,+\lambda}(\xi,\xb,\tau,Q^2,\vk^2) 
K_\nu K_\mu + \cdots\,.
\label{kernel-exp}
\ea
Higher order terms in $K_\mu$ are neglected. In the spirit of the
modified perturbative approach, the $\vk^2$ terms in the propagator
denominators are kept as has been done for the longitudinal
amplitude. As already mentioned the first term in Eq.\ \req{kernel-exp},
generated by $q\sla^\prime\eps\sla_V(\pm 1)$,    
vanishes as a consequence of the number of $\gamma$ matrices in the 
trace and, evidently, the second one as well after integration on $\vk$. 
Hence, the third term in \req{kernel-exp} is the leading one for 
transversely polarized mesons and leads to the subprocess amplitudes
\be 
{\cal H}^{Vi}_{+\lambda,+\lambda}\= -\frac{g_{\perp \mu\nu}}{2}
               \int d\tau\,\frac{d\vk^2}{16\pi^2} \vk^2
                        \Psi_{VT}(\tau,\vk^2)\,
       \Delta {\cal F}^{\,i\,\nu\mu}_{+\lambda,+\lambda}\,,
\label{sub-amp-T}
\ee
in which $g_\perp$ is the transverse metric tensor~\footnote{All its
  elements are zero except $g_\perp^{11}=g_\perp^{22}=-1$.}. 
Note that the wavefunctions for longitudinally and transversely
polarized vector mesons are different in general. The transverse 
amplitude is of order $\vk^2/(M_VQ)$. Noting that 
$\langle\, \vk^2\,\rangle^{1/2}/M_V$ is of order unity, one realizes that 
the transverse amplitude is suppressed by 
\be
{\cal M}_{+\nu',+\nu}\propto \langle \vk^2\rangle^{1/2}/Q\,,
\label{eq:TT-supp}
\ee
with respect to the one for longitudinally polarized vector mesons. 

Working out the kernels, one finds after summation over the parton
helicity that the kernel for transverse photon and meson polarization 
is obtained from the longitudinal one, Eq.\  \req{kernel-long}, by
the replacement
\be
T_s-T_u \longrightarrow \frac{\vk^2}{2}\,\frac{Q}{M_V} \Big[T_s T_a -
T_u T_b\Big]\,.
\label{kernel-trans}
\ee
The new propagator denominators read
\be
T_a\=\frac1{{\tau}Q^2+\vk^2}\,, \qquad T_b\= \frac1{\bar{\tau}
  Q^2+\vk^2}\,.
\ee
We note in passing that $T_s$ and $T_u$ represent the denominators of
the gluon propagators in the LO subprocess $\gamma^* q\to Vq$ while
$T_a$ and $T_b$ belong to the quark propagators (see Fig.\
\ref{fig:feynman}). With the help of partial fractioning 
($i=s, u$, $j=a,b$)
\be 
T_i T_j = \frac1{\vk^2}\,\Big[c_i T_i + c_j T_j\Big]\,,
\ee
we can cast the Fourier transform of the transverse subprocess
amplitude into exactly the same form as for the longitudinal one,
Eq.\ \req{mod-amp}. The kernel is then a linear combination of four
Fourier transformed propagators. 

The denominators of the parton propagators in \req{q-kernel} are
either of the type
\be
   {T}_1 \= \frac{1}{\vk^2 + d_1Q^2}\,,
\ee
or
\be
T_2 \= \frac1{\vk^2 - d_2 (\xb\pm\xi) Q^2 - i \veps}\,.
\label{eq:T2}
\ee
where $d_i\geq 0$. The Fourier transforms of these propagator terms can 
readily be obtained:
\ba
\hat{T}_1 &=& \frac1{2\pi}\, K_0(\sqrt{d_1}\,bQ)\,,\nn\\
\hat{T}_2 &=& \frac{1}{2\pi}\, 
           K_0\left(\sqrt{d_2(\pm\xi -\xb)}\, bQ\right)\;
                    \theta(\pm\xi -\xb) \nn\\
             &+&  \frac{i}{4} \, 
             H^{(1)}_0\left(\,\sqrt{d_2\,(\,\xb\pm\xi\,)}\,bQ\right)\; 
                        \theta(\,\xb\pm\xi\,)\,,
\label{eq:T2-FT}
\ea
where $K_0$ and $H_0^{(1)}$ are the zeroth order modified Bessel
function of second kind and Hankel function, respectively. 
 
\section{The double distribution model}
\label{sec:GPD}
As in Refs.\ \ci{first,second} the GPDs are constructed from the PDFs 
with the help of double distributions \ci{mul94,rad98}. Since this
construction is described in detail in our previous papers we will only
recapitulate a few basic elements of it. For details we refer
to Refs.\ \ci{first,second}. The main advantage of this construction 
is the warranted polynomiality of the resulting GPDs and the correct 
forward limit $\xi, t \to0$. It is well-known that, at low $x$, the  
parton distribution functions behave as powers $\delta_i$ of $x$. 
These powers are assumed to be generated by Regge poles \ci{landshoff71,feynman}. 
We generalize this behavior of the PDFs by assuming that the $t$ 
dependence of the double distributions and hence the GPDs are also
under control of Regge behavior. Linear Regge trajectories are assumed 
for small $-t$ 
\be
 \alpha_i \= \alpha_i(0) + \alpha'_i t\,, \qquad \qquad i=g, {\rm sea}, {\rm val}\,,
\ee 
with $\delta_i=\alpha_i(0)$ for quarks and, as a consequence of the
familiar definition of the gluon GPD which reduces to $\xb g(\xb)$ in 
the forward limit, $\delta_g=\alpha_g(0)-1$ for gluons. The 
trajectories are accompanied by Regge residues assumed to have an 
exponential $t$ dependence with parameters $b_i$. The following 
ansatz for the double distributions associated with the GPDs $H_i$ is
therefore employed (cf.\ \ci{second})
\be
f_i(\beta,\alpha,t)\= {\rm e}^{b_it}\,\mid\beta\mid^{-\alpha_i'\,t}\, h_i(\beta)\,
                   \frac{\Gamma(2n_i+2)}{2^{2n_i+1}\,\Gamma^2(n_i+1)}
                   \,\frac{[(1-|\beta|)^2-\alpha^2]^{n_i}}
                           {(1-|\beta|)^{2n_i+1}}\,,
\label{DD}
\ee
where
\ba
h_g(\beta)\; &=& |\beta|g(|\beta|)\, \hspace*{0.154\textwidth} n_g\;=2\,, \nn\\
h^q_{\rm sea}(\beta) &=& q_{\rm sea}(|\beta|)\;{\rm sign}(\beta)
                         \hspace*{0.07\textwidth} n_{\rm sea}=2\,, \nn\\
h^q_{\rm val}(\beta) &=& q_{\rm val}(\beta)\, \Theta(\beta)    
                   \hspace*{0.117\textwidth} n_{\rm val}=1\,.
\label{function-h}
\ea
For the decomposition of the double distribution into valence and sea
contribution we follow the procedure proposed in Ref.\ \ci{diehl03}
and write 
\ba
f^q_{\rm val}(\beta,\alpha,t) &=&\big[f^q(\beta,\alpha,t)
              + f^q(-\beta,\alpha,t)\big]\, \Theta(\beta)\,, \nn\\
f^q_{\rm sea}(\beta,\alpha,t) &=&f^q(\beta,\alpha,t)\,\Theta(\beta)\,
                -\,f^q(-\beta,\alpha,t)\,\Theta(-\beta)\,.
\ea
In the forward limit, $\xi, t\to 0$, this decomposition is conform to
the usual definition of sea and valence quark PDFs.

The GPDs are related to the double distributions by the integral
\be
H_i(\xb,\xi,t)\=\int_{-1}^1 d\beta\,\int_{-1+|\beta|}^{1-|\beta|} d\alpha\,
                 \delta(\beta+\xi\alpha-\xb)\,f_i(\beta,\alpha,t)\,.
\label{GPD-DD}
\ee
For convenience we employ an expansion of the PDFs ($\beta>0$)
\be
h_i(\beta) \=
\beta^{-\delta_i}\,(1-\beta)^{\,2n_i+1}\;\sum_{j=0}^3\,c_{ij}\,\beta^{j/2}\,,
\label{pdf-exp}
\ee
which is particularly useful at low $\beta$ and allows to perform the
integral \req{GPD-DD} term by term analytically; its use is also
convenient if the integral \req{GPD-DD} is carried out numerically. The factor
$(1-\beta)^{2n_i+1}$ serves for canceling the corresponding factor in Eq.\
\req{DD} and has the additional welcome feature of roughly accounting
for the $\beta\to 1$ behavior of the PDFs. The ansatz \req{pdf-exp}
results in a corresponding expansion of the GPDs 
\be
H_i(\xb,\xi,t) \= {\rm e}^{b_it}\,\sum_{j=0}^3\, c_{ij}\,H_{ij}(\xb,\xi,t)\,.
\label{GPD-exp}
\ee
The definition of the GPDs is completed by the relations
\be
H^g(-\xb,\xi,t) \= H^g(\xb,\xi,t)\,, \qquad H_{\rm
  sea}^q(-\xb,\xi,t)\=-H_{\rm sea}^q(\xb,\xi,t)\,,
\label{symmetry}
\ee
and
\be
H_{\rm val}^q(\xb,\xi,t) \=0\, \hspace*{0.1\textwidth} -1 \leq \xb <
-\xi\,.
\ee
In Eq.\ \req{GPD-DD} so-called $D$ terms for the gluons and the flavor singlet
quark combination are ignored \ci{pol99}. The $D$ terms ensure the 
appearance of the highest powers of the skewness in the moments of the 
GPDs. They only contribute to the less important real parts of the 
amplitudes since their support is the region $-\xi < x < \xi$. The 
corresponding imaginary parts are related to the GPDs at 
$\xb\= \xi (1+2\vk^2/(\tau Q^2))$ which lies outside the support of
the $D$ terms. We take this in vindication of neglecting the $D$ terms.

\begin{figure}[t]
\begin{center}
\includegraphics[width=0.31\textwidth, bb= 147 318 462 720,clip=true]
{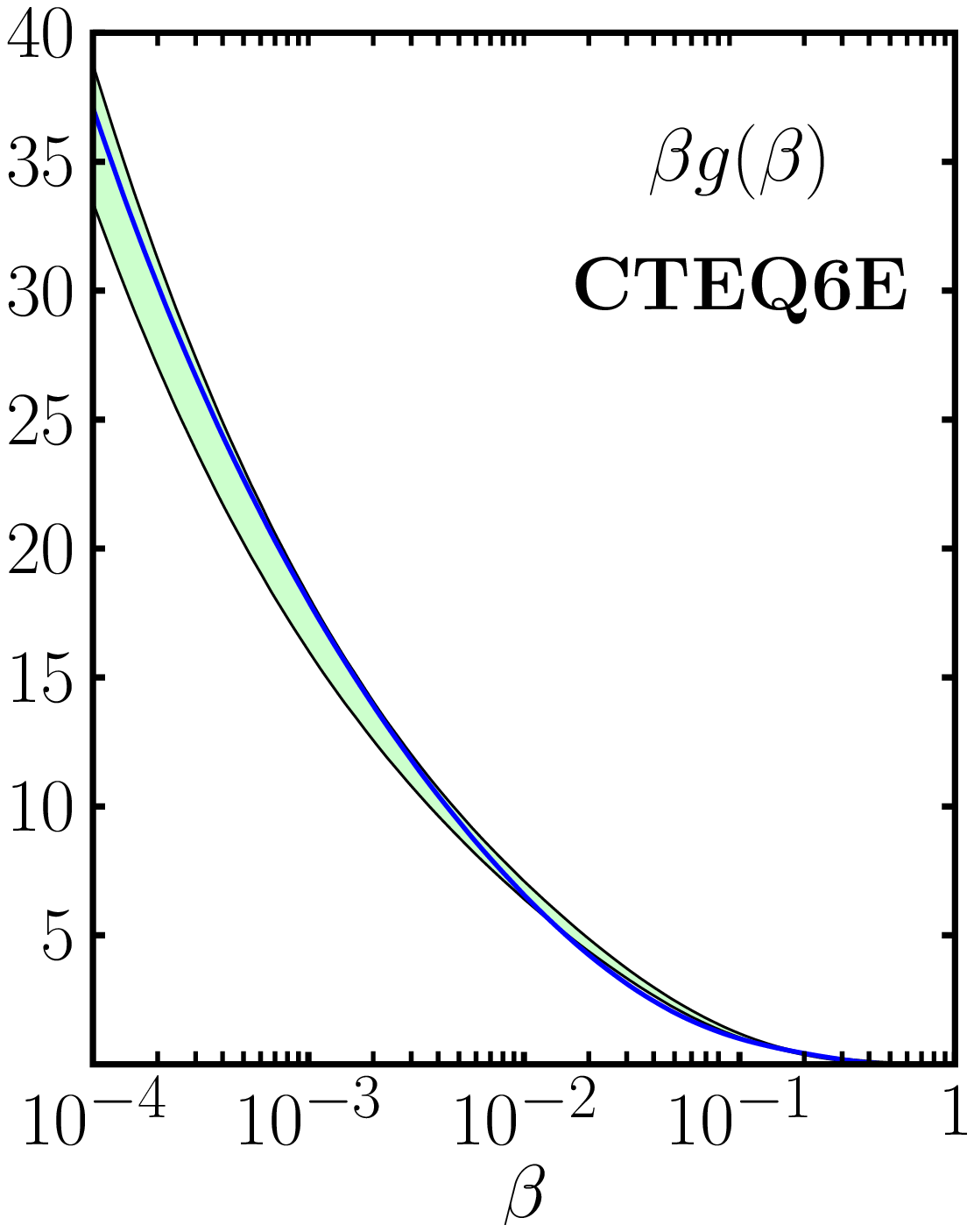}\hspace*{0.1\tw}
\includegraphics[width=0.46\textwidth, bb= 38 348 518 743,clip=true]{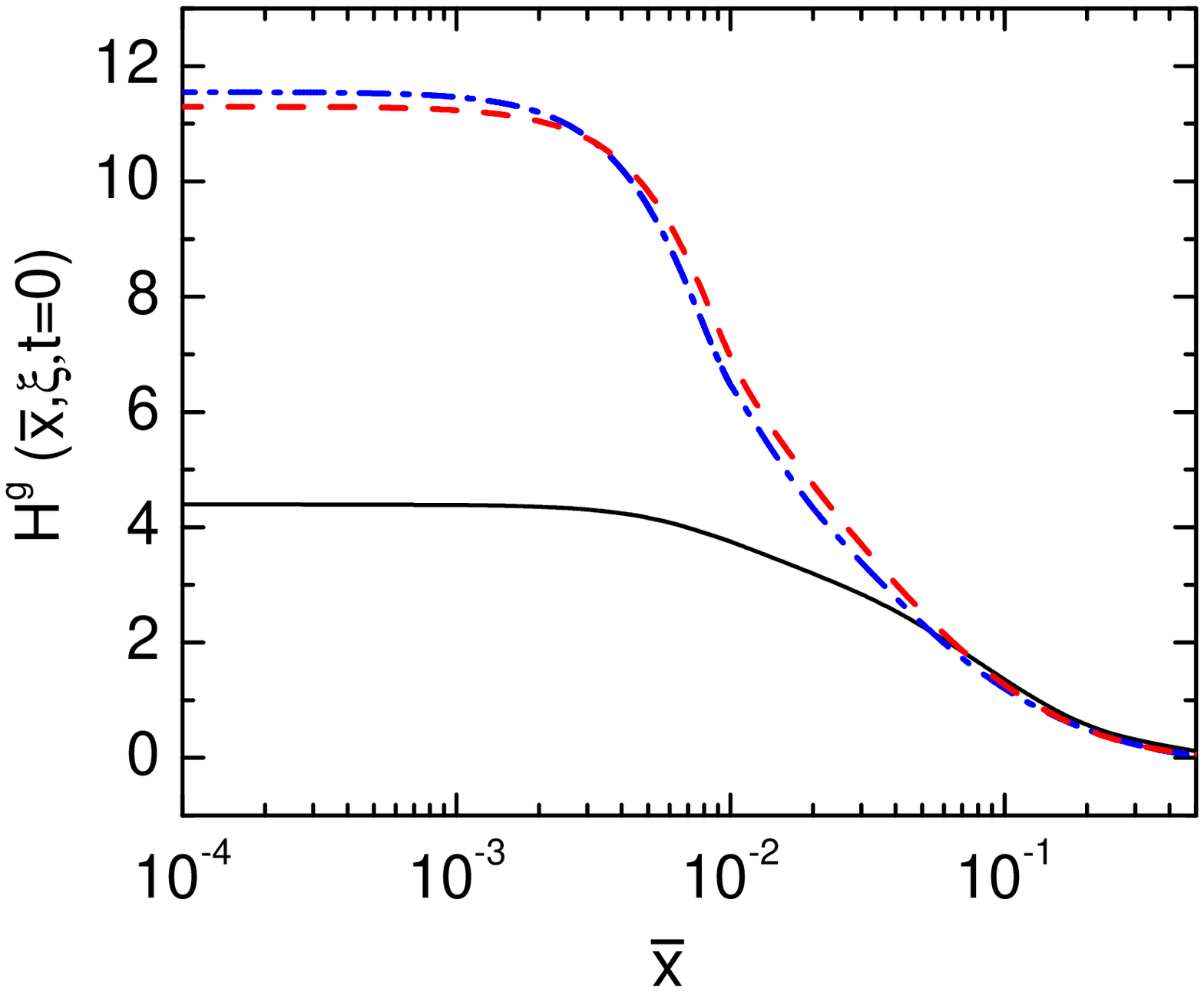}
\caption{Left: The CTEQ6 gluon PDF at $Q^2=100\,\gev^2$ as compared to
  our fit (solid line). The band indicates the Hessian errors of the PDFs. 
Right: The gluon GPD $H^q$ at $\xi=0.01$ and $t=0$. Solid line: $H^g$
  at the initial scale $4\,\gev^2$. Dash-dotted (dashed) line: at the
  scale of $40\,\gev^2$ evolved with the Vinnikov code \ci{vinnikov} 
(approximate evolution).} 
\label{fig:pdf}
\end{center}
\end{figure}

For the expansions of the PDFs we will use the same parameters
as in Ref.\ \ci{second} with the exception of a little change. ZEUS 
\ci{zeus07} now provides data on the cross section for $\rho$
production up to $Q^2=100\,\gev^2$ whereas in \ci{second} the fits to
the CTEQ6M PDFs were made for $Q^2\le 40\,\gev^2$. As one may check,
the CTEQ6M gluon and sea quark PDFs are not well described above $40\,\gev^2$
by the expansion quoted in Ref.\ \ci{second} but the addition of a 
$L^2$-term to $\delta_g$
\be
\delta_g \= 0.10+0.06L-0.0027\,L^2\,,
\label{delta}
\ee
improves the fits to the gluon and sea quark PDFs considerably as can
be seen from Fig. \ref{fig:pdf} ($L=\ln{Q^2/Q^2_0}$, $Q^2_0=4\,\gev^2$). 
The $L^2$ term is irrelevant below $40\,\gev^2$. The valence quarks 
are not needed at high $Q^2$. The parameters of the expansions 
\req{pdf-exp} are quoted in Tab.\ \ref{tab:parH}. We stress that with 
the exception of $\delta_g$, they are identical to those used in Ref.\ 
\ci{second}. In the quoted ranges of $Q^2$ and $\beta$ the fits to the 
PDFs agree very well with the CTEQ6M solution; they are always well 
inside the band of Hessian errors quoted in \ci{cteq6}. Larger values 
of $\beta$ are irrelevant to us since the region $0.5 \lsim \beta$ 
affects the real parts of the amplitudes only marginally; the 
contributions are less than $0.5 \%$. From the fitted PDFs the GPDs are 
evaluated with the help of Eq.\ \req{GPD-DD}. 

In an attempt to keep the GPD model simple we assume
\be
H^u_{\rm sea} \= H^d_{\rm sea} \= \kappa_s\,H^s_{\rm sea}\,,
\label{eq:sea}
\ee
where the flavor symmetry breaking factor is parameterized as
\be
\kappa_s \= 1+ 0.68/(1+0.52\,\ln{Q^2/Q^2_0})\,,
\label{eq:kappas}
\ee
as obtained from a fit to the CTEQ6M PDFs. 

\begin{table*}
\renewcommand{\arraystretch}{1.4} 
\begin{center}
\begin{tabular}{|c|| c |c | c | c|}
\hline
         & gluon & strange & $u_{\rm val}$ & $d_{\rm val}$ \\
\hline
$\delta$ & Eq.\ \req{delta} & $1+\delta_g$ & 0.48 & 0.48 \\ 
$\alpha'$& $0.15\,\gev^{-2}$& $0.15\,\gev^{-2}$&$0.9\,\gev^{-2}$&$0.9\,\gev^{-2}$\\ 
$c_0$    & $2.23+0.362\,L$ & $\phantom{-}0.123+0.0003\,L$
         & $1.52+0.248\,L$& $ 0.76+0.248\,L$ \\
$c_1$    & $\phantom{-}5.43-7.00\,L$ & $-0.327-0.004\,L$ & 
                                $2.88-0.940\,L$ & $\phantom{-}3.11-1.36\,L$ \\
$c_2$    & $-34.0+22.5\,L$ & $\phantom{-}0.692-0.068\,L$& $-0.095\,L$
         & $-3.99+1.15\,L$ \\
$c_3$    & $\phantom{-}40.6 -21.6\,L$ & $-0.486+0.038\,L$ & $0$ & $0$ \\
\hline
\end{tabular}
\end{center}
\caption{The parameters appearing in the expansion \req{pdf-exp} of 
the PDFs ($L=\ln{Q^2/Q^2_0}$, $Q^2_0=4\,\gev^2$).} 
\label{tab:parH}
\renewcommand{\arraystretch}{1.0}   
\end{table*} 

As in Ref.\ \ci{second} we take for the slope of the gluon trajectory
the value $\alpha_g'=0.15\,\gev^{-2}$. Since the sea quark PDFs are mainly
driven by evolution for $Q^2\gsim 4\,\gev^2$ it is assumed that 
$\alpha_{\rm sea}(t)=\alpha_g(t)$. A standard trajectory is adopted
for the valence quark Regge pole - $\alpha_{\rm val}(t)=0.48+0.9\,\gev^{-2}\,t$. 
The parameter of the gluon residue is fixed from a fit against the HERA
data for $\rho$ \ci{h1} and $\phi$ production \ci{zeus05}:
\be
b_g \= b_{\rm sea} \= 2.58\, \gev^{-2} + 0.25\,\gev^{-2}\,
  \ln{\frac{m^2}{Q^2+m^2}}\,,
\label{constant-slope}
\ee 
The $\rho$ and $\phi$ slopes of the cross sections practically fall
together at HERA energies; there are only minor differences at low $Q^2$.
The parameter $b_g$ given in \req{constant-slope} leads to a $t$
dependence of the differential cross section in perfect agreement with
the recent ZEUS data \ci{zeus07}. The parameter of the valence quark
residue is taken to be zero. This is in accord with the findings of
the nucleon form factor analysis proposed in Ref.\ \ci{DFJK4} in which 
the zero-skewness GPDs have been determined.

We emphasize that the evolution of the GPDs is taken into account by
us only approximately through the evolution of the PDFs. This is
reasonable since at low $\xi$ the imaginary part of the gluon (and
sea quark) contribution dominates which is $\propto H^g(\xi,\xi,t)$
and therefore approximately equals $2\xi g(2\xi)$ at low $\xi$ (see, 
for instance \ci{first}). Its real part as well as the valence
quark contributions are only of importance near $4\,\gev^2$, the
initial value of the evolution. The very time-consuming numerical 
integration on $x, {\bf b}$ and $\tau$ forces us to use this 
approximative treatment of the evolution. In order to demonstrate the 
quality of our approximation we compare in Fig.\ \ref{fig:pdf} the 
gluon GPD at $40\,\gev^2$ either obtained from our approximation or from 
evolving $H^g$ from the initial scale of $4\,\gev^2$ using the evolution 
code developed by Vinnikov \ci{vinnikov}. Only minor differences are
to be noticed.

\section{Results on cross sections}
\label{sec:trans}
Before we present our results obtained from the handbag approach we
have to specify the meson wave functions used in the evaluation of the
amplitudes. As in Refs.\ \ci{first,second} we will take Gaussian 
wave functions ($j=L,T$) 
\be
\Psi_{Vj}(\tau,\vk) \= 8\pi^2\sqrt{2N_c} f_{Vj}(\mu_F) a_{Vj}^2 
                 \Big[1+B_2^{Vj}(\mu_F) C_2^{3/2}(2\tau-1)\Big]\,
                   {\rm exp}{[-a_{Vj}^2\vk^2/(\tau\bar{\tau})]}\,.
\label{wavefunction}
\ee
Transverse momentum integration leads to the meson distribution
amplitudes for which we allow for the second Gegenbauer moment besides
the asymptotic form. The meson decay constants for longitudinally
polarized vector mesons are known from the electronic decays of the
vector mesons while those for transversely polarized mesons are taken
from QCD sum rules \ci{braun96}. In contrast to the decay constants 
$f_{VL}$ the latter ones are scale dependent
\be
f_{VT}(\mu_F) \= f_{VT}(\mu_0)\,
\left(\frac{\als(\mu_F)}{\als(\mu_0)}\right)^{4/27}\,.
\ee
Note that the decay constants for transversely polarized vector mesons 
always appear in the combination $f_{VT}/M_V$, i.e. there is only one
independent parameter in practice. In fact we use a typical constituent 
mass of $300\,\mev$ for $M_V$ and the QCD sum rule result \ci{braun96}  
$f_{VT}/f_{VL}\simeq 0.8$ at the scale of $\mu_0=1\,\gev$.
The Gegenbauer coefficients $B_2^{VL}$ have been found to be zero in
the analysis of the longitudinal cross section \ci{second}. Those for
the transverse case are fitted to the data on the $\sigma_T$ or the
cross section ratio $R=\sigma_L/\sigma_T$. The Gegenbauer coefficients 
are scale dependent
\be
B_2^{Vj}(\mu_F) \= B_2^{Vj}(\mu_0)\,
   \left(\frac{\als(\mu_F)}{\als(\mu_0)}\right)^{\gamma_{2j}}\,,
\ee
where $\gamma_{2L}=50/81$ and $\gamma_{2T}=40/81$ \ci{shifman}. 
Finally, the transverse size parameters $a_{Vj}$ in 
\req{wavefunction} are either fitted to $\sigma_L$ or to $\sigma_T$
depending on the polarization of the vector meson. The values for the
various parameters are compiled in Tab.\ \ref{tab:parameters}. Those for
the longitudinal case are identical to the parameters used in
\ci{second}.
\begin{table*}[tb]
\renewcommand{\arraystretch}{1.4} 
\begin{center}
\begin{tabular}{|c|| c |c | c| c|}
\hline
parameter  & $\rho_L$ & $\rho_T$ & $\phi_L$ & $\phi_T$ \\
\hline
$f_V [\mev]$      & 209    &  167   & 221    &  177   \\
$a_V [\gev^{-1}]$ &  0.75  &  1.0   & 0.70   &  0.95   \\
$B_2^V$           & 0.0    &  0.10   & 0.0    &  0.10  \\ 
\hline
\end{tabular}
\end{center}
\caption{The parameters appearing in the wavefunction \req{wavefunction}, 
quoted at the scale $\mu_0=1\,\gev$.}
\label{tab:parameters}
\renewcommand{\arraystretch}{1.0}   
\end{table*} 

The assessment of the theoretical uncertainties deserves special
considerations. The results on cross sections (and other observables) 
are subject to parametric errors. The main uncertainties stem from the
Hessian errors of the set of CTEQ6 PDFs. Since for the longitudinal 
cross section the parameters of the corresponding wave functions are 
adjusted such that good agreement between the data on $\sigma_L$ and 
the handbag results is achieved, there is no substantial additional 
uncertainties from the longitudinal \wf. Results for $\sigma_L$
evaluated from sets of PDFs other than CTEQ6 also fall into 
the error bands in most cases (an exception is set for instance by the
PDFs determined in Ref.\ \ci{GRV,GRV2}) provided these PDFs are treated
in analogy to the CTEQ6M set, i.e.\ they are fitted to the 
expansion \req{pdf-exp} by forcing them to behave Regge-like with 
powers $\delta_i$ as described above, and, if necessary, readjusting 
the transverse size parameters. The uncertainties in the ratio, $R$, 
of the longitudinal and transverse cross sections mainly arise from 
the uncertainties of the wavefunctions for transversely polarized
vector mesons (i.e.\ from the Gegenbauer coefficients, the transverse 
size parameters and from the ratio $f_{VT}/M_V$). The errors due to 
those of the GPDs or PDFs which are mainly responsible for the 
theoretical uncertainties of the cross sections, cancel to a large 
extent in the ratio. 

\begin{figure}[p]
\begin{center}
\includegraphics[width=0.45\textwidth, bb= 39 350 526 727,clip=true]
{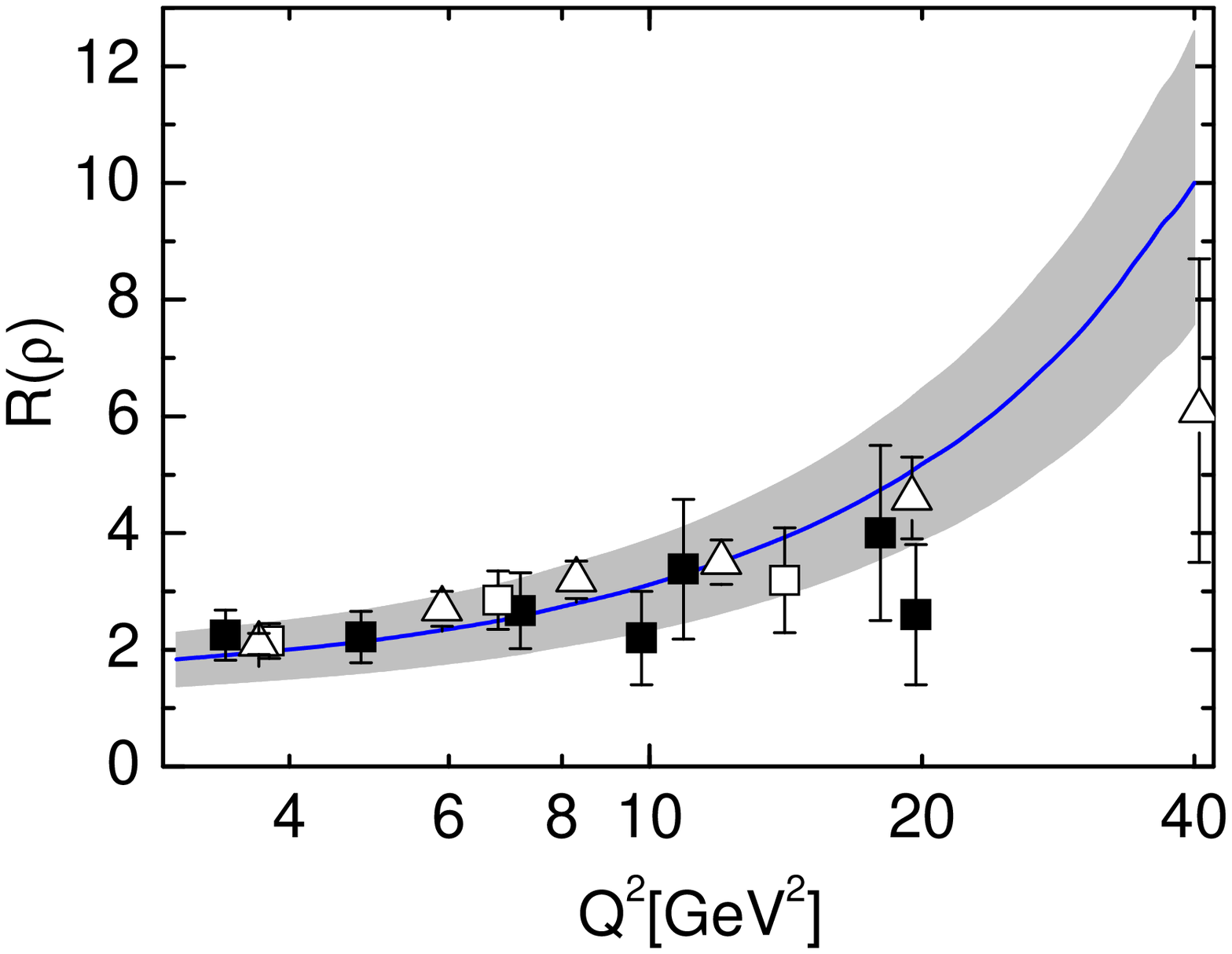}\hspace*{0.1\tw}
\includegraphics[width=0.46\textwidth, bb= 46 320 550 700,clip=true]
{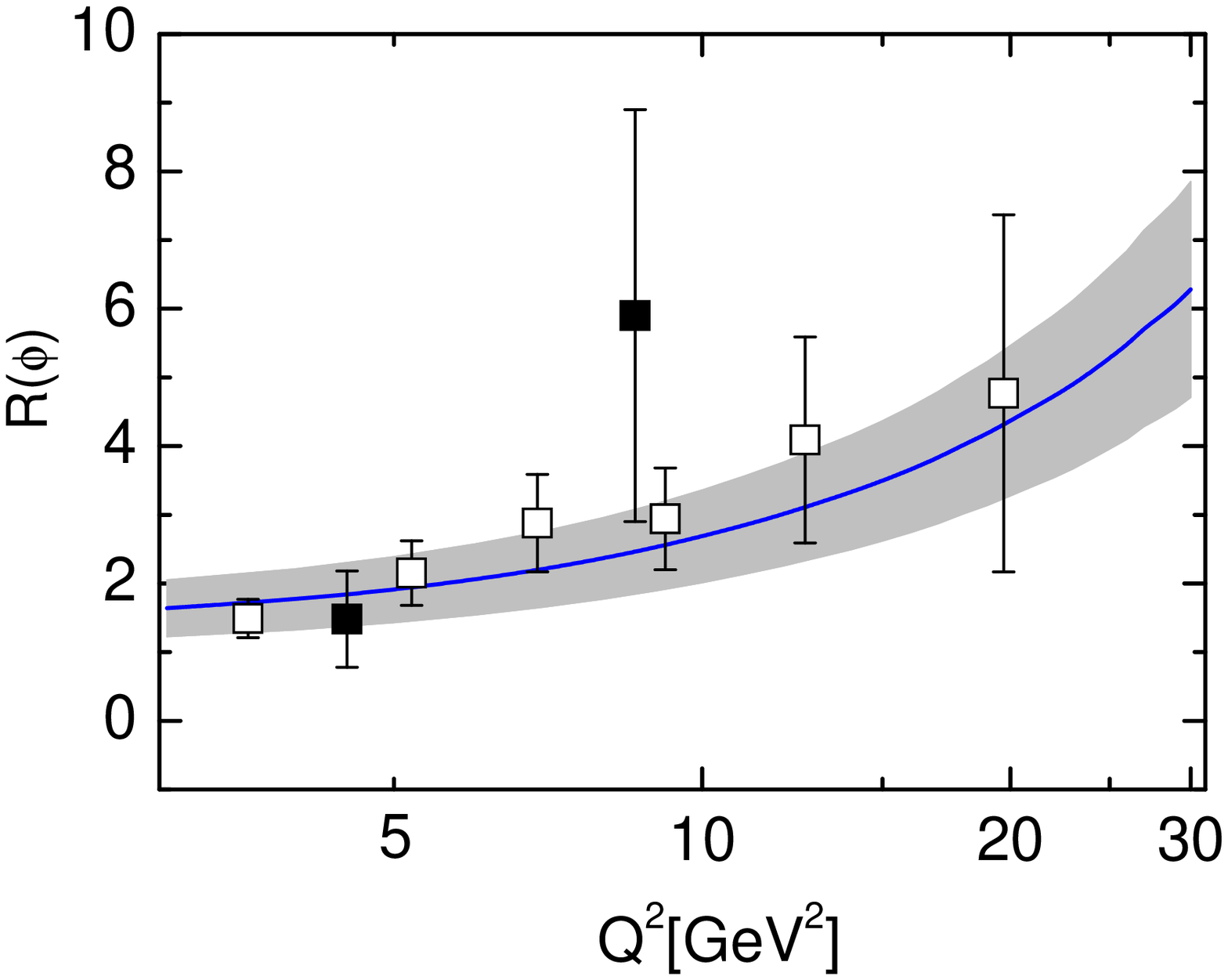}
\caption{The ratio of longitudinal and transverse cross sections
 for $\rho$ (left) and $\phi$ (right) production versus $Q^2$ at
 $W= 90\,\gev$ and $75\,\gev$, respectively.
 Data taken from H1 \protect\ci{h1,adloff,h1-96} (solid
 squares) and ZEUS \protect\ci{zeus05,zeus98} (open squares),
 respectively. The recent ZEUS data \ci{zeus07} are shown as open
 triangles. The solid lines represent the handbag results with the
 shaded bands indicating the uncertainties of the predictions.}
\label{fig:ratio1}
\vspace*{0.05\tw}
\includegraphics[width=0.41\textwidth, bb= 53 321 540 738,clip=true]
{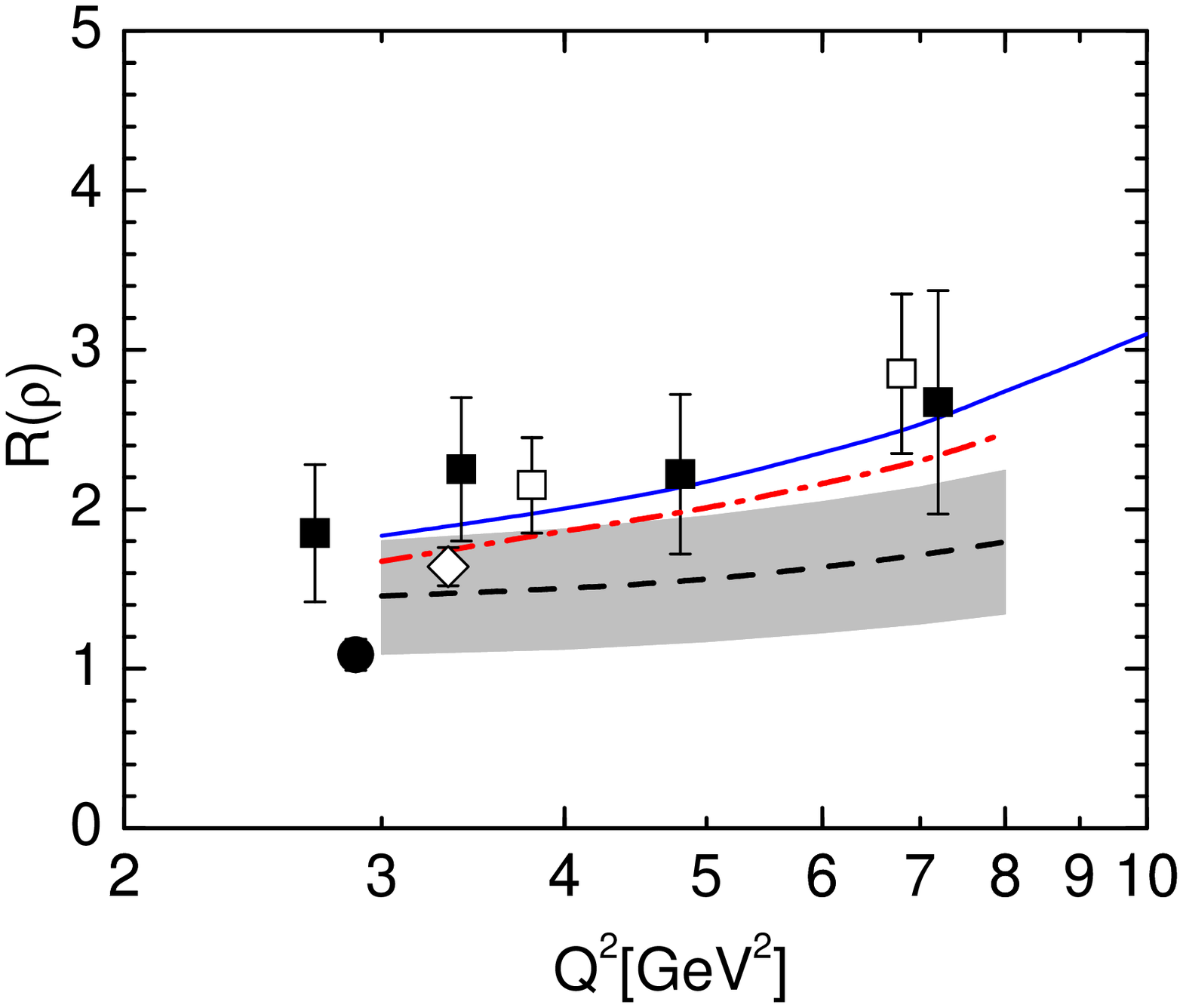}\hspace*{0.1\tw}
\includegraphics[width=0.45\textwidth, bb= 54 321 553 698,clip=true]
{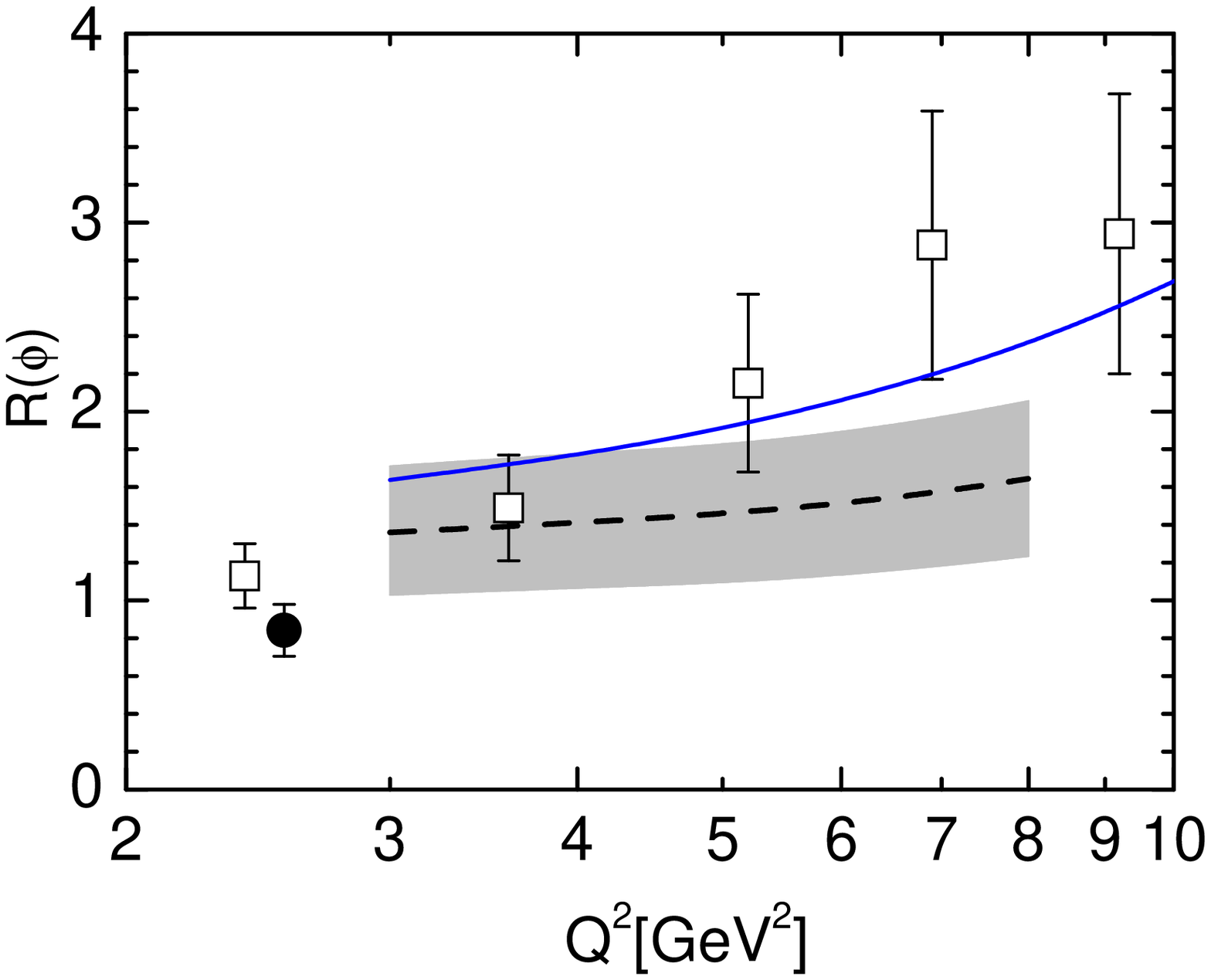}
\caption{Handbag results for $R$ on $\rho$ (left) and $\phi$ (right)
  production at  $W=75 (10,5)\,\gev$  shown as solid (dash-dotted, 
  dashed) line. Data taken from Refs.\ \ci{h1,zeus05,zeus98}. 
  Preliminary data from HERMES \ci{hermes-draft,hermes-phi} (solid
  circle) and  COMPASS \ci{compass} (diamond). The error bands 
  are only shown at $W=5\,\gev$. For further notation refer to Fig.\ 
  \ref{fig:ratio1}.}
\label{fig:ratio2}
\end{center}
\end{figure}

The results for the longitudinal cross sections are the same as in Ref.\
\ci{second}. We refrain from showing them again. The experimental data 
on the cross section ratio $R$ are customarily determined from the 
measured SDME $r_{00}^{04}$ by the relation
\be
R\=\frac{\sigma_L}{\sigma_T} \= \frac1{\varepsilon}
\,\frac{r_{00}^{04}}{1-r_{00}^{04}}\,.
\label{eq:R}
\ee 
The SDME in \req{eq:R} is understood to be integrated over the full 
range of $t$ available in a given experiment~\footnote{
The cross sections $\sigma_L$ and $\sigma_T$ have been obtained from 
integrating the differential cross sections over that range of $t$.}.
The theoretical and experimental results on the ratio $R$ are compared
in Figs.\ \ref{fig:ratio1} and \ref{fig:ratio2}. In general we achieve 
very good agreement with experiment in particular with regard to the 
theoretical uncertainties displayed as shaded bands in the plots. 

The ratio $R$ is mildly energy dependent for $W$ larger than about
$10\,\gev$ while, for lower energies, it exhibits a somewhat stronger
energy dependence, in particular at larger values of $Q^2$, see Fig.\
\ref{fig:ratio2}. This implies differences in the energy dependences
of $\sigma_L$ and $\sigma_T$ which can be traced back to the different
hard scattering kernels (see e.g.\ Eq.\ \req{kernel-trans}) and the
varying \wf s for longitudinally and transversely polarized vector 
mesons, in particular to the different values of the r.m.s.\ $\vk$ 
($\sim 1/a_V$). An energy dependence of $R$ seems to be indicated by 
the preliminary HERMES \ci{hermes-draft} and COMPASS \ci{compass} data 
although confirmation of this observation is demanded.  Most of the older
experiments have rather large errors such that a definite conclusion
on a possible $W$ dependence cannot be drawn at present.  
Unfortunately, for $\phi$ production there is only one low energy data 
point available \ci{hermes-phi} and this point is measured  at the very 
low value of $Q^2=2.6\,\gev^2$ which lies outside the range where the 
handbag approach, in its present form, can be trusted. We note that for 
$\phi$ production the energy dependence of $R$ is even milder than for 
the case of the $\rho$; the results for $W=10\,\gev$ practically fall 
together with those at $75\,\gev$.
 
\begin{figure}[pht]
\begin{center}
\includegraphics[width=.45\textwidth, bb=20 323 552 744,clip=true]
{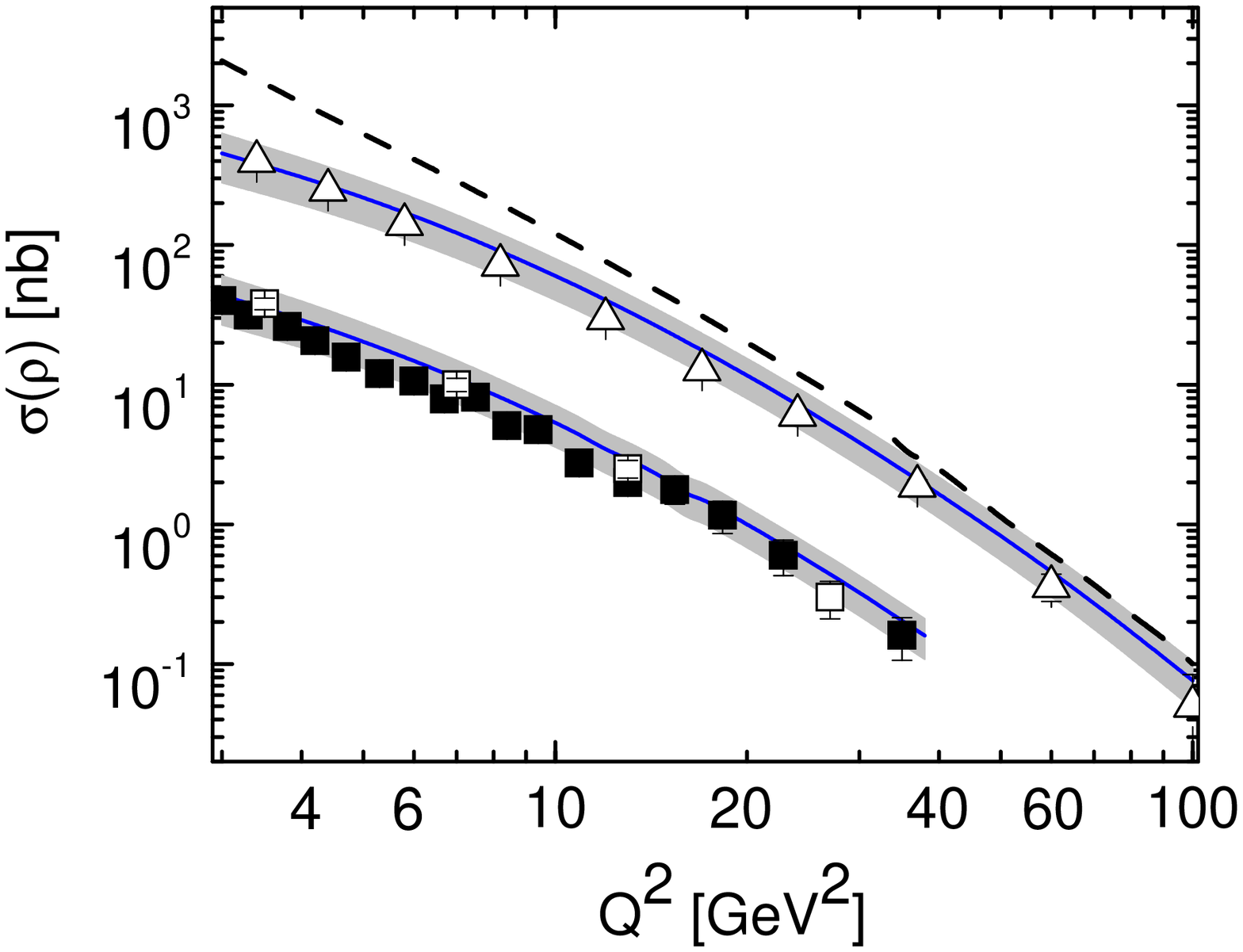}\hspace*{1.0em}
\includegraphics[width=.44\textwidth,bb= 25 316 563 744,clip=true]
{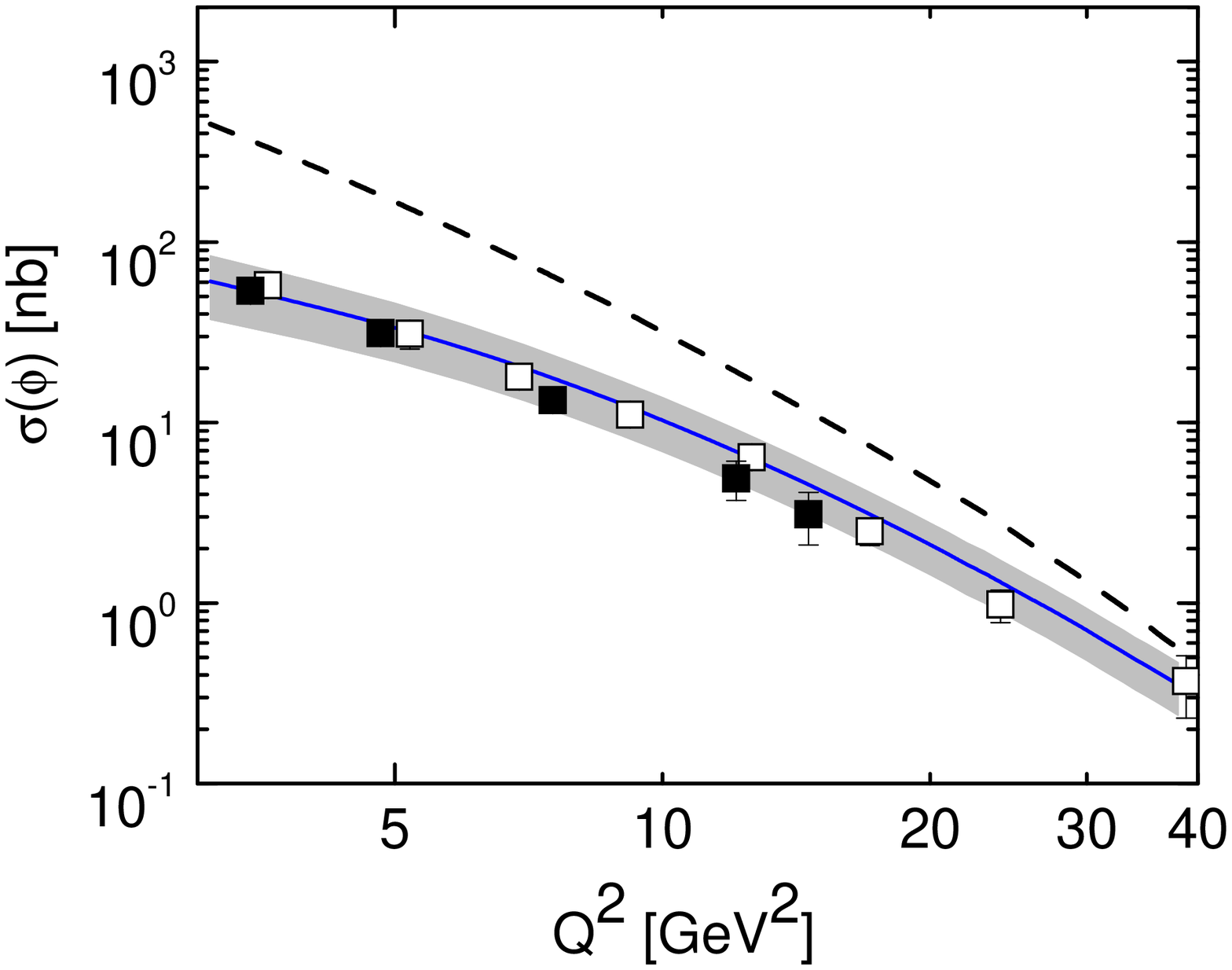}
\caption{The $Q^2$-dependence of the cross sections for $\rho$ (left)
  and $\phi$ (right) production at $W=90$ ($\rho$) and $75\,\gev$ 
  ($\rho, \phi$). Data are taken from ZEUS \ci{zeus07,zeus05,zeus98}
  and H1 \ci{h1,adloff}. For $\rho$ production at $W=75\,\gev$ the 
  data and theoretical results are divided by 10 for the ease of 
  legibility. The dashed lines represent the leading-twist contribution. 
  For further notations refer to Fig.\ \ref{fig:ratio1}.}
\label{fig:sigma-Q}
\vspace*{0.05\tw}
\includegraphics[width=.35\textwidth, bb=37 255 506 756,clip=true]
{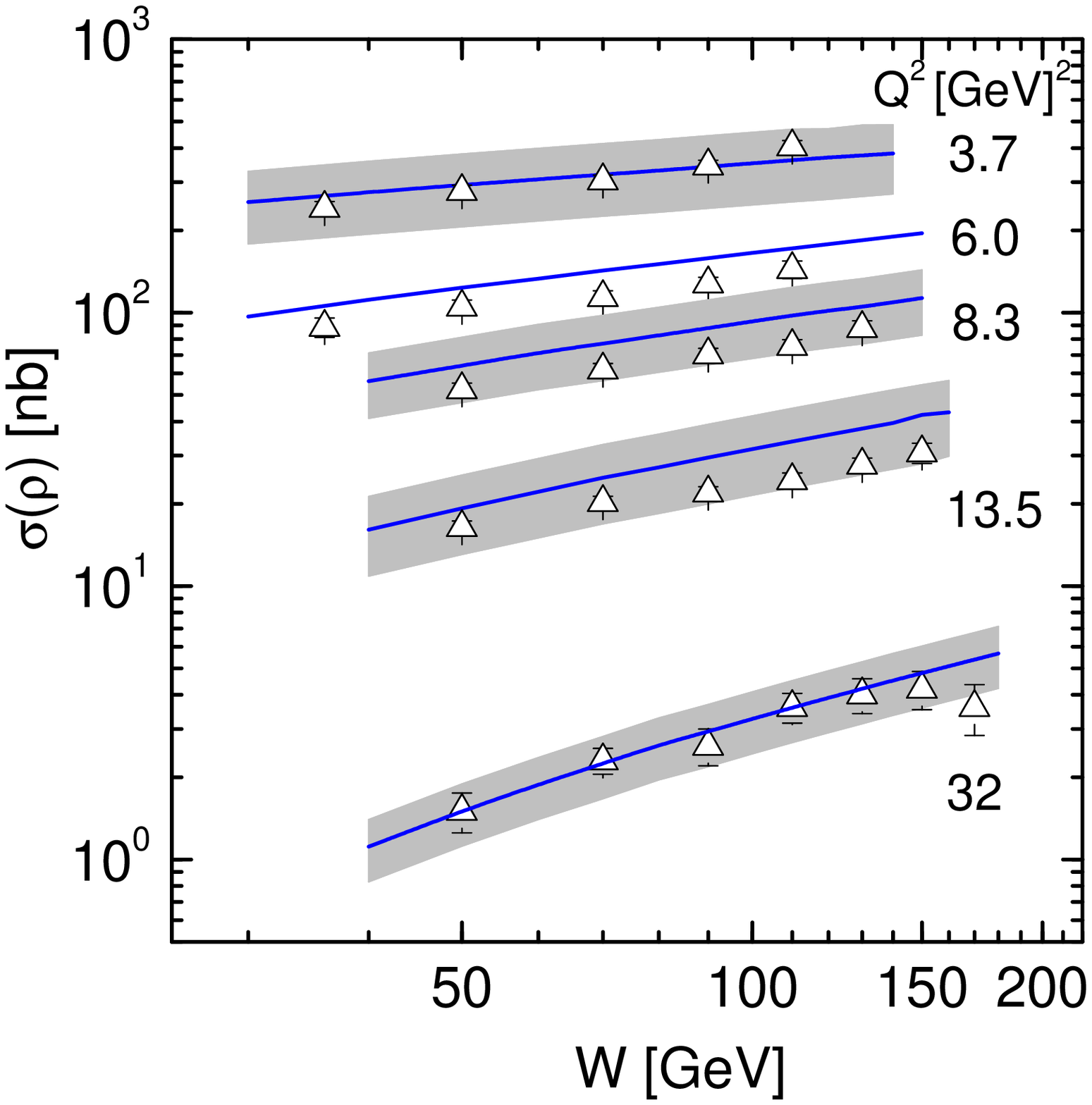}\hspace*{0.05\tw}
\includegraphics[width=.46\tw,bb=8 319 523 742,%
clip=true]{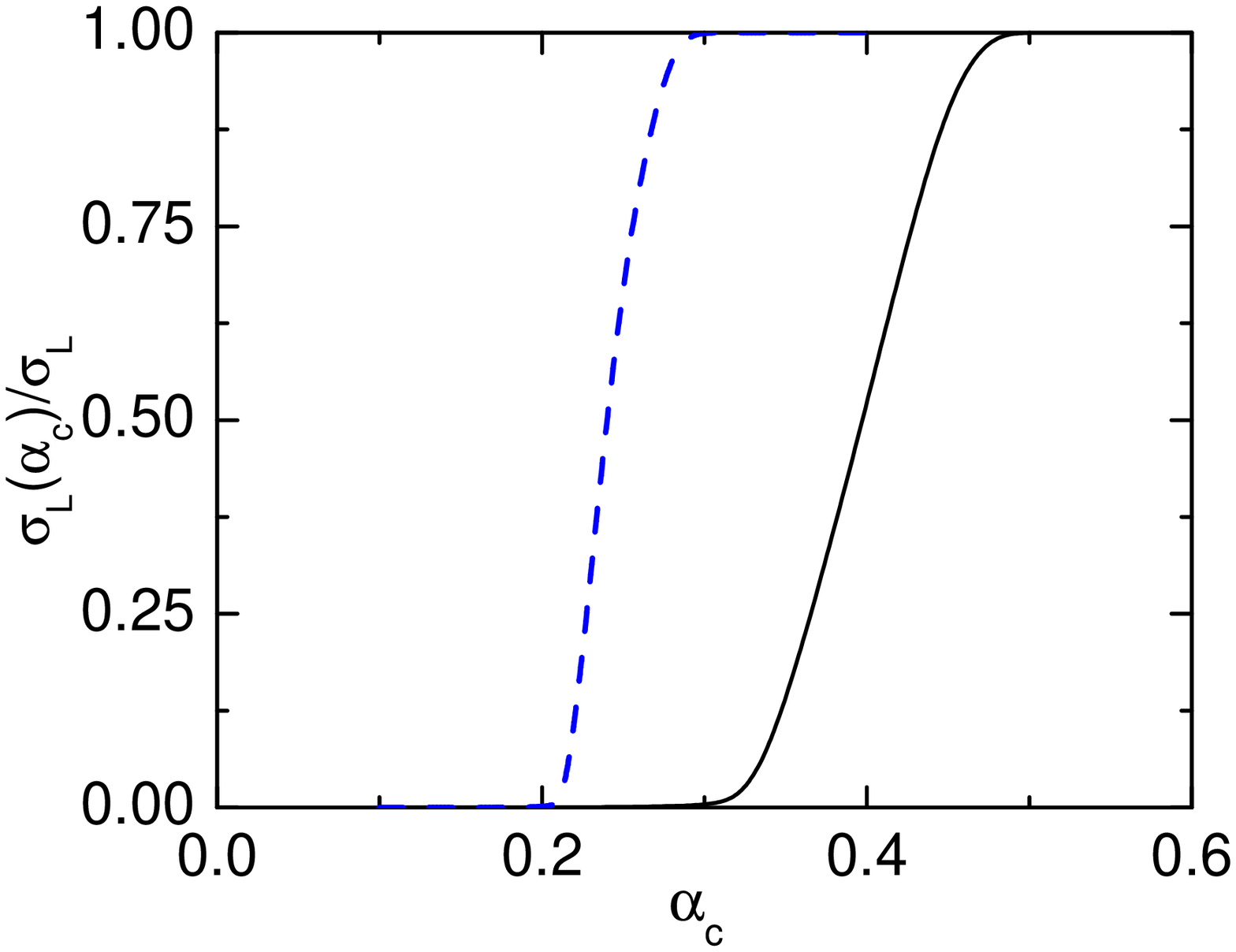}
\caption{Left: The cross section for $\rho$ production versus $W$ at a
set of $Q^2$ values. The data are taken from ZEUS \ci{zeus07}. For 
further notations refer to Fig.\ \ref{fig:ratio1}. The error band at 
$6\,\gev^2$ is omitted for the ease of legibility. Right: The 
accumulation profile of the amplitude in dependence of $\alpha_c^{\rm crit}$ 
at $Q^2=4$ (solid line) and $40\,\gev^2$ (dashed line) at $W=75\,\gev$.}
\label{fig:sigma-W}
\end{center}
\end{figure}

In Fig.\ \ref{fig:sigma-Q} the handbag results on the $\rho$ and $\phi$
cross sections are compared to experiment. Again good agreement with
the H1 \ci{h1,adloff} and ZEUS \ci{zeus07,zeus05,zeus98} data is to be 
observed in a large range of $Q^2$. The leading-twist contribution to
these cross sections, i.e.\ $\sigma_L$ evaluated in collinear
approximation, is also shown. Although the leading-twist contribution 
approaches the experimental cross section with increasing $Q^2$ there 
is still a small difference of about $1.5\sigma$ between both at 
$Q^2=100\,\gev^2$ for $\rho$ production. Note that even at that 
value of $Q^2$ the transverse cross section which is included in the 
full one and also represents a power correction to the leading-twist
result, is not negligible; it amounts to about $10\%$. The
leading-twist contribution is about $20\%$ larger than the one
obtained within the modified perturbative approach at
$Q^2=100\,\gev^2$, i.e.\ the corrections due the quark transverse
momentum have not yet disappeared completely. In Fig.\ 
\ref{fig:sigma-W} the energy dependence of the $\rho$ cross section at 
a set of $Q^2$ values is displayed. Within errors agreement is to be 
seen with the ZEUS data \ci{zeus07}. This is not a surprise since the 
power $\delta_g$, related to the intercept of the gluonic Regge
intercept, is fixed by the energy dependence of the HERA data on the 
cross sections, see Ref.\ \ci{second}. As $\sigma_L$, see \ci{second}, 
the full cross section at HERA energies is dominated by the gluon 
contribution although the sea quarks are not negligible.
Including the interference with the gluon  the sea quarks contribute
about $25\%$ in the case of the $\phi$ and $40\%$ in the case of the
$\rho$ at $Q^2=4\,\gev^2$. The larger sea-quark contribution in the
latter case is due to the flavor-symmetry breaking factor $\kappa_s$
\req{eq:kappas}. Flavor symmetry breaking in the sea is important
for the ratio of the $\phi$ and $\rho$ cross sections, see Ref.\ 
\ci{second}. Neglecting the sea quarks or assuming a flavor symmetric 
sea leads to an incorrect $\phi-\rho$ ratio. Going to energies lower 
than about $10\,\gev$ the valence quark contributions gradually become 
perceptible for $\rho$ production. At, say, $5\,\gev$ the valence
quarks are responsible for about $40\%$ of the cross section. We
stress that the three contributions have to be added coherently.  
There are substantial interference terms which increase the cross 
sections markedly. For instance, at $W=5\,\gev$ and $Q^2=4\,\gev^2$ 
the $\rho$ cross section is doubled by the interference terms.

A remark concerning the $t$ dependence of the differential cross
sections is in order. As we pointed out in \ci{second} the cross 
sections drop exponentially  with $t$ to a very good approximation. 
Their slopes are approximately given by $2b_g$ (see
\req{constant-slope}) plus a contribution from the 
gluonic Regge trajectory (see Eq.\ \req{DD} and Ref.\ \ci{second}).  
At large $W$ the slopes of the differential cross sections for 
longitudinally and transversely polarized photons are the same 
while, at low $W$, the valence quarks generate small differences. 
Consequently, the ratio $R$ is nearly $t$ independent a fact that is
in agreement with the ZEUS measurement \ci{zeus07}. 
 
Finally, we check the theoretical consistency of the modified
perturbative approach. Consistency is meant in the sense that the bulk
of the perturbative contributions should be accumulated in regions
where the strong coupling $\als$ is sufficiently small. To find out
whether or not this is the case, we set the  integrand in  
\req{mod-amp} equal to zero in those regions where 
$\als(\mu_R)>\alpha_s^{\rm crit}$, and evaluate, say, the longitudinal 
cross section for $\rho$ production as a function of 
$\alpha_s^{\rm crit}$. Consulting Fig.\ \ref{fig:sigma-W} where the
accumulation profile is shown, one observes that almost the entire
result is accumulated in a comparatively narrow region of $\als$
around 0.4 at $Q^2=4\,\gev^2$ and 0.25 at $40\,\gev^2$. Thus the
effective renormalization scales in these two cases are about 1.5 and 
$15\, \gev^2$, respectively. Hence, our results on cross sections are 
theoretically self-consistent with regard to the above-mentioned 
criterion. Contributions from the end-point regions where the momentum 
fraction $\tau$ tends either to zero or to one and where, in collinear 
approximation, the renormalization scale becomes very small, are 
sufficiently suppressed. 

\section{Results on spin density matrix elements}
\label{sec:sdme}
In a number of experiments \ci{zeus07,h1,adloff,compass,zeus00}
the SDMEs have been extracted from the measured decay angular
distributions of the vector mesons. The formalism of the SDMEs, i.e.\
their relations to the amplitudes for the process $\gamma^* p \to Vp$,
has been developed by Schilling and Wolf \ci{schilling} long-time ago.
This work has recently been repeated and extended to the case of a
transversely polarized target proton by Diehl \ci{diehl07}. Since in
the experimental papers the notation of Ref.\ \ci{schilling} is used
throughout, we will adhere to it here in order to facilitate
comparison.

The SDMEs are given by properly normalized bilinear combinations of 
$\gamma^* p \to Vp$ amplitudes. Due to the symmetry relations
\req{sym-N} and \req{sym-U} there are no interference terms between
${\cal M}^N$ and ${\cal M}^U$ in the case of unpolarized
electroproduction of vector mesons. The normalizations read
\ba
N_L&=& 2\,\sum_{\nu^\prime}\,
         \mid  {\cal M}^N_{0\nu^\prime,0+}\mid^2\,, \nn\\
N_T&=& 2\, \sum_{\nu^\prime} \left[ 
       \mid  {\cal M}^N_{+\nu^\prime,++}\mid^2\,
     +  \mid  {\cal M}^U_{+\nu^\prime,++}\mid^2\, \right]\,,
\label{norm}
\ea
where helicity-flip $\gamma^*\to V$ transitions are neglected as we do
throughout this work. Up to a phase space factor, see Eq.\ \req{sigma}, 
the normalizations are the differential cross sections for
longitudinally and transversely polarized vector mesons. Since the  
SDMEs also provide a possibility to study the role of $\widetilde{H}$ 
we do not ignore the contributions from the corresponding amplitude 
${\cal M}^U_{+\nu',++}$ in \req{norm} for later reference.
  
Obviously, the SDMEs are functions of $Q^2, t$ and $W$. Due to 
limitations in statistics it is not possible to measure the
SDMEs as a functions of the three variables. Thus, frequently the
SDMEs are presented as functions of one variable for average values
of the other variables. For instance the SDMEs are quoted as a 
function of $Q^2$ for an average value of $W$ and integrated over all 
$t$ available in a given experiment. Only in this case, and this is an 
important one to which we will mainly refer to in the following, the 
normalizations $N_L$ and $N_T$ refer to the respective integrated cross 
sections $\sigma_L$ and $\sigma_T$ up to the phase-space factor and 
eventually neglected suppressed amplitudes. If for instance the SDMEs
are presented as a function of $t$ the normalization represents 
differential cross sections at a fixed value of $t$ but integrated
over certain ranges of $Q^2$ and $W$. 

\begin{figure}[p]
\begin{center}
\includegraphics[width=0.74\textwidth, bb= 111 403 560 720,clip=true]
{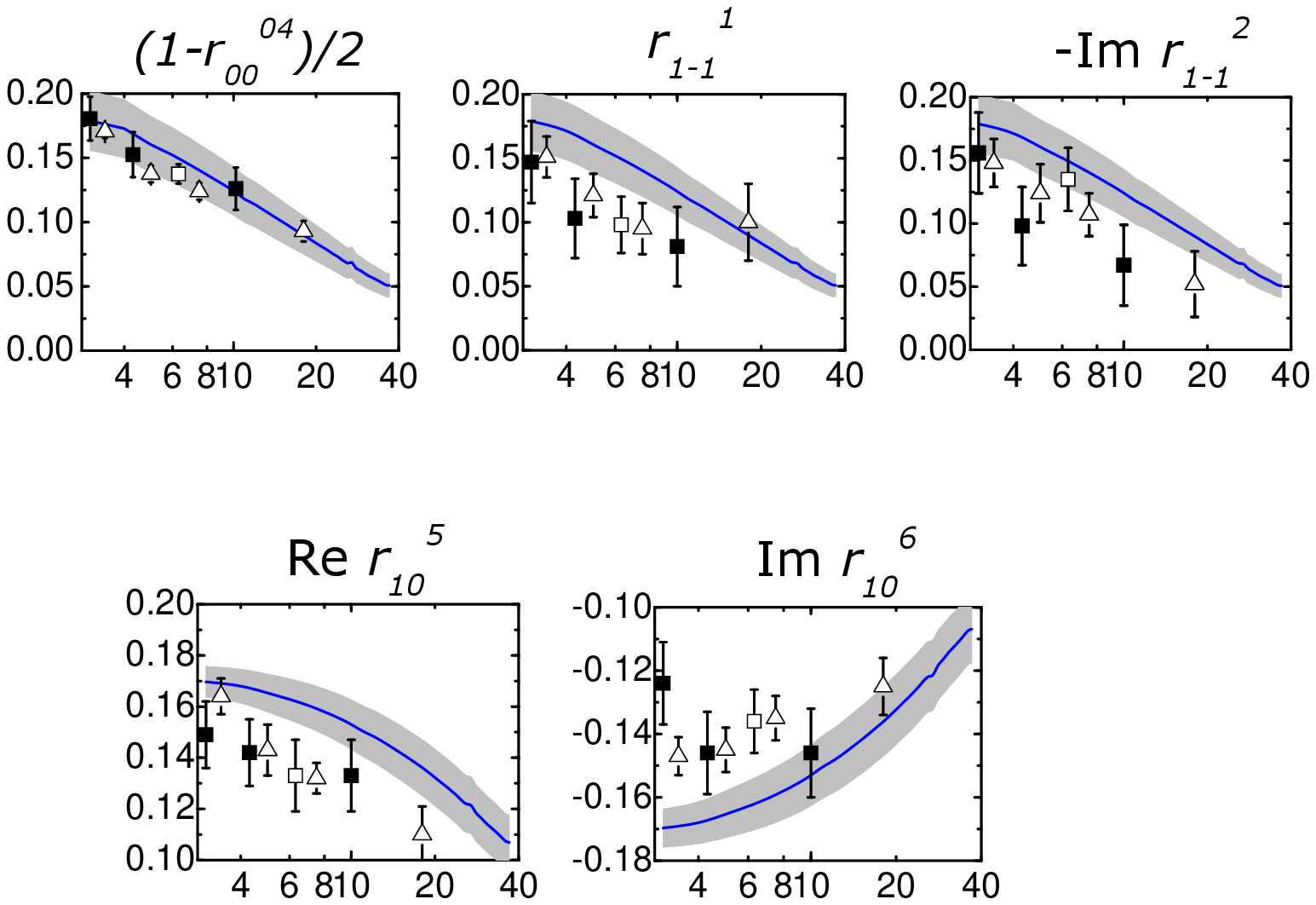}
\caption{The SDMEs for the $\rho$ meson versus $Q^2$ (in $\gev^2$) at 
 $W=75\,\gev$. Data taken from \ci{zeus07,h1,zeus00}. Note the
 different scales at the axis of ordinates. For further notations
 refer to Fig.\ \ref{fig:ratio1}.} 
\label{fig:sdme-rho-hera}
\vspace*{0.027\tw}
\includegraphics[width=0.74\textwidth, bb= 111 418 560 717,clip=true]
{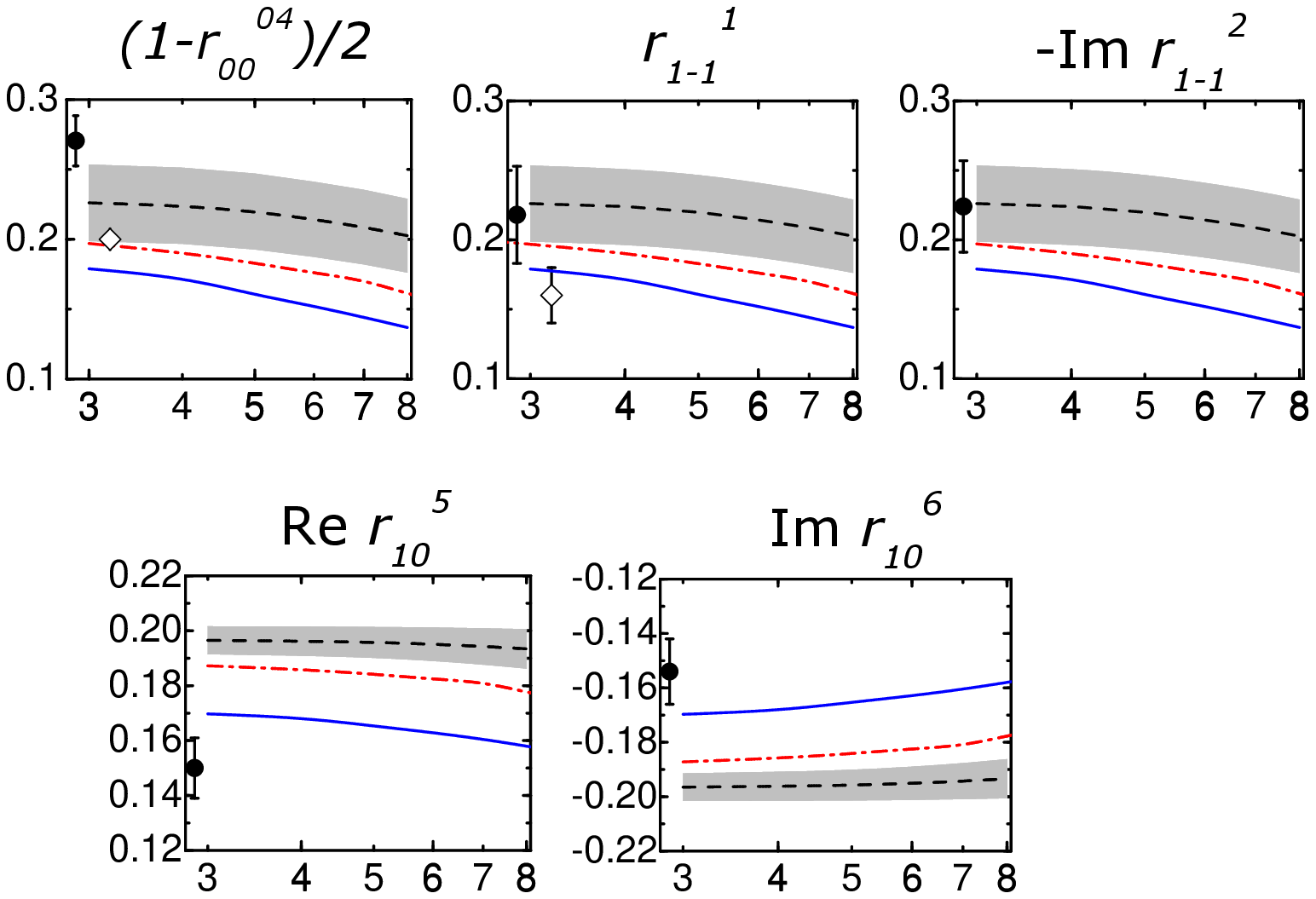}
\caption{SDMEs for the $\rho$ meson versus $Q^2$ (in $\gev^2$) at
  $W=5, 10$ and $75\,\gev$ (dashed, dash-dotted and solid line,
  respectively). Preliminary data taken from HERMES
  \ci{hermes-draft} (solid circles) and COMPASS \ci{compass}
  (diamonds). Error bands only shown for $W=5\,\gev$. For further
  notations refer to Figs.\ \ref{fig:ratio1} and \ref{fig:ratio2}.}
\label{fig:sdme-rho-hermes}
\end{center}
\end{figure}

\begin{figure}[p]
\begin{center}
\includegraphics[width=0.74\textwidth, bb= 111 422 560 718,clip=true]
{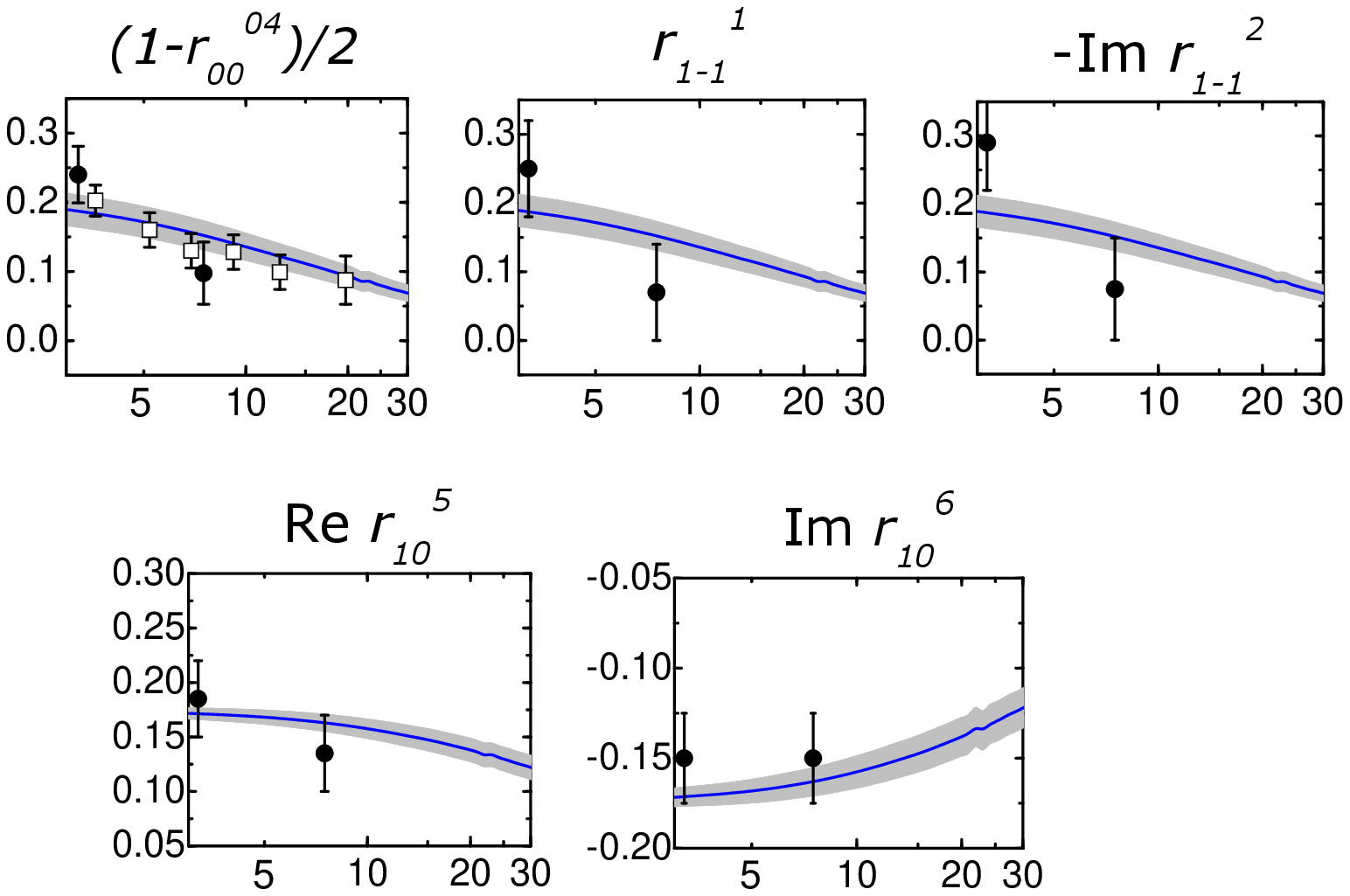}
\caption{SDMEs for the $\phi$ meson versus $Q^2$ (in $\gev^2$) at
    $W=75\,\gev$.  Data taken from \ci{zeus05,adloff}. For further
    notations refer to Fig.\ \ref{fig:ratio1}. } 
\label{fig:sdme-phi-hera}
\vspace*{0.20\tw}
\includegraphics[width=0.74\textwidth, bb= 113 570 565 718,clip=true]
{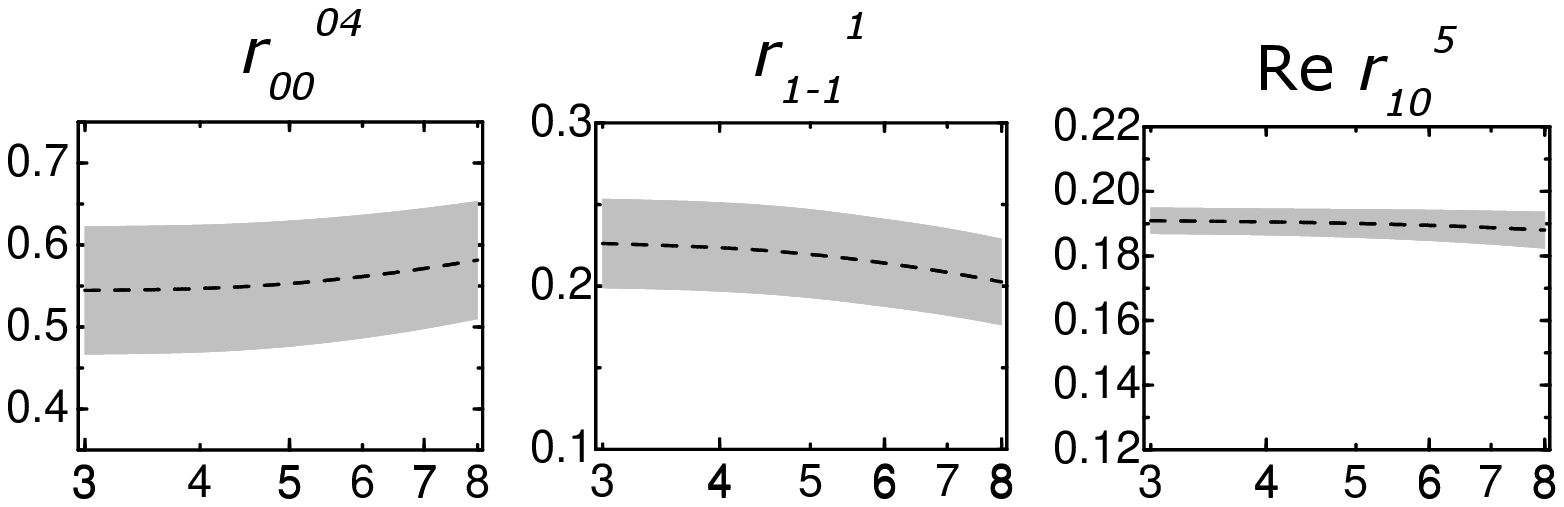}
\caption{SDMEs for the $\phi$ meson versus $Q^2$ (in $\gev^2$) at
    $W=5\,\gev$.}
\label{fig:sdme-phi-hermes}
\end{center}
\end{figure}

The SDMEs allow for a separation of the absolute values of the two
amplitudes, ${\cal M}^N_{0+,0+}$ and ${\cal M}^N_{++,++}$, and for a
test of their relative phases. Whether or not it is justified to
neglect the other amplitudes can be examined as well. The SDMEs 
$r_{00}^{04}$, $r_{1-1}^1$ and  ${\rm Im}\,r_{1-1}^2$ are related to 
the cross section ratio $R$. In terms of the amplitudes we are 
investigating these SDMEs read
\ba
r_{00}^{04} &=& \frac{2\veps}{N_T+\veps N_L}\,\sum_{\nu^\prime} \mid {\cal
    M}^N_{0\nu^\prime,0+}\mid^2\,, \nn\\
r_{1-1}^1 &=&- {\rm Im} r_{1-1}^2 \=  \frac{1}{N_T+\veps N_L}\,
            \sum_{\nu^\prime} \left[
           \mid {\cal M}^N_{+\nu^\prime,++}\mid^2
           -\mid {\cal M}^U_{+\nu^\prime,++}\mid^2\right]\,.
\label{eq:sdme}
\ea
Neglecting the amplitude ${\cal M}^U$ too these expressions simplify to
\be
1-r_{00}^{04} \= 2 r_{1-1}^1 \= -2 {\rm Im}\, r_{1-1}^2 \=
\frac1{1+\veps R}\,.
\ee 
In Figs.\ \ref{fig:sdme-rho-hera}, \ref{fig:sdme-rho-hermes} and 
\ref{fig:sdme-phi-hera} this prediction of our handbag approach is
compared to experiment. The agreement is in general good within errors 
although with occasional exceptions. Thus, for $\rho$ production, 
$r_{1-1}^1$ and ${\rm Im}\, r_{1-1}^2$ are not perfectly reproduced. 
On the other hand, their sum, measuring the double flip transitions 
$\gamma^*_T\to V_{-T}$, is nicely in agreement with zero. A further 
check is provided by the $t$ dependence of the SDMEs. Since in the
handbag approach the $t$ dependence solely comes from the GPDs and 
these are identical for the two amplitudes the above SDMEs should be 
nearly flat in $t$ which is indeed the case experimentally within errors
\ci{zeus07,h1}. Hence, the above three SDMEs are consistent with our 
assumptions and do not provide a significant  signal for helicity-flip 
amplitudes or contributions from $\widetilde{H}$. The $W$ dependence 
of the handbag predictions is displayed in Fig.\ \ref{fig:sdme-rho-hermes}. 
It is evidently very mild. In particular the results for the SDMEs at 
$W=75$ and $90\,\gev$ fall practically together.

After having checked that the absolute values of our two amplitudes
are in fair agreement with experiment, we now turn to their relative phase
$\delta_{LT}$. In terms of our two amplitudes the SDMEs relevant for
the verification of the phase, read
\ba
{\rm Re}\, r_{10}^5 &=& - {\rm Im}\, r_{10}^6 \= \frac1{\sqrt{2}}\,
\frac1{N_T+\veps N_L}\, {\rm Re}\, \left[{\cal M}^N_{++,++} {\cal
    M}^{N*}_{0+,0+}\right]\nn\\
  &=& \frac1{2\sqrt{2}}\, \frac{\sqrt{R}}{1+\veps R}\,
\cos{\delta_{LT}}\,.
\label{deltaLT}
\ea

Predictions for the SDMEs in the case of $\phi$ production are shown 
in Figs.\ \ref{fig:sdme-phi-hera} and \ref{fig:sdme-phi-hermes}. 
Fair agreement with experiments can be seen. For $\rho$ production,
on the other hand, a conflict is to be noted, see Figs.\
\ref{fig:sdme-rho-hera} and \ref{fig:sdme-rho-hermes}. The data
\ci{zeus07,h1,hermes-draft} require a rather large phase although with
strong fluctuations ($10-30^\circ$) while the handbag approach
provides only a small value for it (e.g.\ $\delta_{LT}=3.1^\circ$ at 
$W=5\,\gev$ and $Q^2=3\,\gev^2$). Whether our model for the 
$\gamma^*_T\to V_T$ amplitude which represents a power correction to 
the leading $\gamma^*_L\to V_L$ amplitude, is inadequate for this
detail needs further investigation. However, that the sum 
${\rm Re}\, r_{10}^5+{\rm Im}\, r_{10}^6$ amounts to only $1\%$ of the 
corresponding difference of these SDME makes it clear that the neglected 
helicity flip $\gamma^*\to V$ transitions in \req{deltaLT} are not
responsible for the observed conflict.

\begin{figure}[ht]
\begin{center}
\vspace*{0.05\tw}
\includegraphics[width=0.31\textwidth, bb= 67 277 541 697,clip=true]
{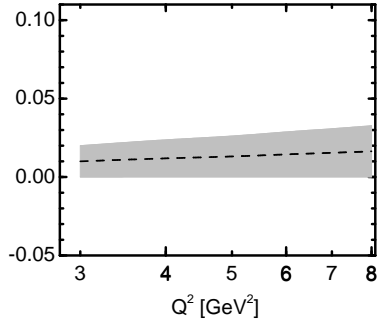}
\caption{The SDMEs ${\rm Im}\, r_{10}^7$ and  ${\rm Re}\, r_{10}^8$ 
for $\rho$ production versus $Q^2$ at $W=5\,\gev$.  
For further notations refer to Fig.\ \ref{fig:ratio1}.}
\label{fig:sdme-rho-hermes-pol}
\end{center}
\end{figure}

The HERMES collaboration has also measured the decay angular
distribution of the $\rho$ in the case of a longitudinally polarized
lepton beam. These measurements which are not yet analyzed, will
provide data on other SDMEs \ci{schilling}. For instance, the SDMEs 
${\rm Im}\, r_{10}^7$ and ${\rm Re}\, r_{10}^8$ also measure the phase 
$\delta_{LT}$ introduced in \req{deltaLT}
\be
{\rm Im}\, r_{10}^7\= {\rm Re}\, r_{10}^8 \= 
\frac1{2\sqrt{2}}\, \frac{\sqrt{R}}{1+\veps R}\, \sin{\delta_{LT}}\,.
\label{pol-sdme}
\ee
Predictions for these SDMEs are shown in Figs.\
\ref{fig:sdme-rho-hermes-pol}. The other polarized SDMEs are only
sensitive to the suppressed amplitudes, see Tab.\ \ref{tab:sdme}. 

In principle also the helicity flip $\gamma^*\to V$ transitions can be
calculated in the proposed handbag approach as well. But these transitions
are strongly suppressed with respect to the amplitude ${\cal
  M}_{0+,0+}$ as the amplitude for $\gamma^*_T\to V_T$ transitions, see
\req{eq:TT-supp}. In fact,
\ba
\gamma^*_T \to V_L &\quad& \propto \quad \frac{\sqrt{-t}}{Q}\, \nn\\ 
\gamma^*_L \to V_T &\quad& \propto \quad
\frac{\sqrt{-t}\,\langle\,\vk^2\rangle^{1/2}}{Q^2}\,\nn\\
\gamma^*_T \to V_{-T} &\quad& \propto \quad
\frac{-t\,\langle\,\vk^2\rangle^{1/2}}{Q^3}\,,
\label{eq:supp}
\ea
where the powers of $\sqrt{-t}/Q$ follow from angular momentum
conservation and the factors $\langle\,\vk^2\rangle^{1/2}/Q$ from the
treatment of transversely polarized vector meson, see \req{eq:TT-supp}.  
The suppression factors given in \req{eq:supp} engender the behavior
of the SDMEs for helicity flip $\gamma^*\to V$ transitions listed in 
Tab.\ \ref{tab:sdme}. The most important helicity-flip amplitude is 
the one for $\gamma^*_T\to V_L$ transitions. It is most clearly seen 
in $r_{00}^5$ as an interference term with the dominant longitudinal 
amplitude. It is definitely non-zero, of the order of $0.1$ for $\rho$ 
and $\phi$ production in experiment \ci{zeus07,h1,hermes-draft}. At 
least HERMES \ci{hermes-draft,marianski} observes a $t$ dependence for it in 
agreement with expectation ($\propto \sqrt{-t}$). The
contribution of the $\gamma^*_T\to V_L$ amplitude to the cross
sections, $R$ and $r_{00}^{04}$ is at the percent level and can
safely be neglected. The other SDMEs related to $\gamma^*_T\to V_L$ 
transitions are experimentally compatible with zero within errors. The 
only exception is to be seen in the recent high statistics ZEUS 
data \ci{zeus07} where ${\rm Im}\, r_{10}^2$ and ${\rm Re}\, r_{10}^1$ 
have very small but non-zero for values of $Q^2$ less than $10\, \gev^2$. 
It would be interesting to see whether their $t$ dependencies are in
agreement with expectation. For the remaining SDMEs being related to 
$\gamma^*_L\to V_T$ and $\gamma^*_T\to V_{-T}$ transitions there is no 
significant deviation from zero at large $Q^2$ experimentally. In view 
of these remarks it is fair to conclude that the neglect of the 
helicity flip $\gamma^*\to V$ transitions in the cross sections for 
$Q^2\geq 3\,\gev^2$ is justified. 

\begin{table*}
\renewcommand{\arraystretch}{1.6} 
\begin{center}
\begin{tabular}{|c|| c |c |}
\hline
SDME         & amplitudes & order \\ 
\hline
$r_{00}^5$, $r_{00}^8$   & $(\gamma^*_T\to V_L)(\gamma^*_L\to V_L)$ & $\sqrt{-t}/Q$ \\
${\rm Re}\,r_{10}^{04}$, ${\rm Re}\,r_{10}^1$, ${\rm Im}\, r_{10}^2$,  
${\rm Im}\, r_{10}^3$ & 
 $(\gamma^*_T\to V_L)(\gamma^*_T\to V_T)$ &
$\sqrt{-t}\langle\,\vk^2\rangle^{1/2}/Q^2$ \\
$r_{00}^1$ & $\mid (\gamma^*_T\to V_L)\mid^2$ & $-t/Q^2$ \\
$r_{11}^5, r_{1-1}^5, {\rm Im}\, r_{1-1}^6$ &  
$(\gamma^*_L\to V_{T})(\gamma^*_T\to V_T)$ &
  $\sqrt{-t}\langle\,\vk^2\rangle/Q^3$  \\
$r_{11}^8, r_{1-1}^8, {\rm Im}\, r_{1-1}^7$ & $(\gamma^*_L\to
V_{T})(\gamma^*_T\to V_T)$ &  $\sqrt{-t}\langle\,\vk^2\rangle/Q^3$ \\  
$r_{1-1}^{04}$, $r_{11}^1$, ${\rm Im}\, r_{1-1}^3$ & 
$(\gamma^*_T\to V_{-T})(\gamma^*_T\to V_T)$ &
 $-t\langle\,\vk^2\rangle/Q^4$ \\
\hline
\end{tabular}
\end{center}
\caption{SDMEs controlled by helicity flip $\gamma^*\to V$ transitions.}
\label{tab:sdme}
\renewcommand{\arraystretch}{1.0}   
\end{table*}

\section{The role of $\widetilde{H}$}
\label{sec:Htilde}
The expression for the amplitude ${\cal M}^U$ is the same as that for 
${\cal M}^N$ given in \req{H-amp} except that the sum of the 
subprocess amplitudes is to be replaced by their difference
\be
{\cal M}_{\mu +, \mu +}^{Ui}(V) \= \frac{e}{2} 
             \sum_a e_a{\cal C}_V^{\,a}\,\int_{\bar{x}_i}^1\,
      d\xb \, \Big[{\cal H}^{Vi}_{\mu+,\mu+}-{\cal H}^{Vi}_{\mu-,\mu-}\Big]
             \,\widetilde{H}^i(\xb,\xi,t)\,.
\label{Ht-amp}
\ee
The superposition of the various quark and gluon contributions is
identical to that for the amplitude ${\cal M}^N$. The unnatural parity
amplitudes satisfy the symmetry relation \req{sym-U}. It is evident 
from \req{Ht-amp} that parity conservation leads to a vanishing
longitudinal amplitude ${\cal M}^U_{0+,0 +}$. The transverse
subprocess amplitude is the same as in \req{sub-amp-T} and 
\req{kernel-trans} except that in the latter equation a plus sign
occurs between $T_sT_a$ and $T_u T_b$ now. 
  
The GPDs $\widetilde{H}$ are again modeled by the double distribution
ansatz \req{DD}. With regard to the symmetry properties of
$\widetilde{H}$ the functions $\widetilde{h}_i$
now take the form
\ba
\tilde{h}_g(\beta)\; &=& |\beta|\Delta g(|\beta|)\;{\rm sign}(\beta)\,,\nn\\
\tilde{h}^q_{\rm sea}(\beta) &=& \Delta q_{\rm sea}(|\beta|)\,,\nn\\
\tilde{h}^q_{\rm val}(\beta) &=& \Delta q_{\rm val}(\beta)\, \Theta(\beta) \,.   
\label{function-ht}
\ea
For the powers $n_i$ in the double distribution ansatz \req{DD} the
same values are taken as for the GPDs $H^i$, see \req{function-h}.  
The decomposition of the double distribution into valence and sea
contributions is made by \ci{diehl03} 
\ba
\tilde{f}^q_{\rm val}(\beta,\alpha,t) &=&\big[\tilde{f}^q(\beta,\alpha,t)
              - \tilde{f}^q(-\beta,\alpha,t)\big]\, \Theta(\beta)\,, \nn\\
\tilde{f}^q_{\rm sea}(\beta,\alpha,t) &=&\tilde{f}^q(\beta,\alpha,t)\,\Theta(\beta)\,
                +\,\tilde{f}^q(-\beta,\alpha,t)\,\Theta(-\beta)\,.
\ea
The double distribution ansatz for $\widetilde{H}^g$ is incomplete
because in moments of this GPD the highest power of $\xi$ are lacking
which leads to difficulties with the analytic properties of the
amplitudes \ci{diehl-ivanov}. We ignore this problem here since 
the contributions from $\widetilde{H}^g$ seem to be unimportant.

The required polarized parton distributions are taken from Ref.\
\ci{BB} and expanded according to 
\be
\tilde{h}_i(\beta) \=
\beta^{-\tilde{\delta}_i}\,(1-\beta)^{\,2n_i+1}\;\sum_{j=0}^3\,
          \tilde{c}_{ij}\,\beta^{j}\,,
\label{pol-pdf-exp}
\ee
using only integer powers. The resulting 
expansion parameters $\tilde{c}_i$ and $\tilde{\delta}_i$ are quoted 
in Tab.\ \ref{tab:par-tilde}. It is expected that the $a_1$ Regge 
trajectory controls the low-$x$ behavior of the polarized valence 
quark PDFs. Since there are no recurrences of the $a_1(1260)$ we are 
forced to assume the standard value of $0.9\,\gev^{-2}$ for the slope  
of the trajectory~\footnote{
Accepting exchange degeneracy for the $a_1$ and $\eta_2(1617)$
trajectories the slope of the trajectory is fixed by the meson
spectrum and is indeed $0.9\,\gev^{-2}$.}.
Combining this with the spin of the $a_1$, we obtain 
$\alpha_{a_1}(0)\simeq -0.36$ for the intercept. Such a low value is
however in conflict with the small-$x$ behavior~\footnote{
Whether this  is a consequence of lack of low-$x$ data in the PDF
analysis or due to disregarded high-lying Regge cuts, is unknown at 
present.}
of the polarized valence quark PDFs determined in Ref.\ \ci{BB}, for
which the power is rather about 0.7. As a compromise we therefore take
the standard value of 0.48 for it. For the slope of the Regge trajectory 
we again take the value of $0.9\,\gev^{_2}$ and for its 
residue $\tilde{b}_{\rm val}=0$. 

The GPDs $\widetilde{H}^i$ are obtained from the functions $f_i$ by
the integral \req{GPD-DD}. They satisfy the relations
\be
\widetilde{H}^g(-\xb,\xi,t) \= -\widetilde{H}^g(\xb,\xi,t)\,, \qquad 
\widetilde{H}_{\rm sea}^q(-\xb,\xi,t)\=\widetilde{H}_{\rm sea}^q(\xb,\xi,t)\,,
\label{symmetry-Ht}
\ee
and
\be
\widetilde{H}_{\rm val}^q(\xb,\xi,t) \=0\, \hspace*{0.1\textwidth} -1 \leq \xb <
-\xi\,.
\ee
We checked that our proposed GPDs $\widetilde{H}_{\rm val}$ are in
agreement with the data on the axial form factor for $-t\lsim
0.6\,\gev^2$ and with the low $-t$ (low $x$) behavior of $\widetilde{H}$ 
determined in \ci{DFJK4}. In constrast to the situation for $H$,
$\widetilde{H}^u_{\rm val}$ and $\widetilde{H}^d_{\rm val}$ have
opposite signs as a consequence of the behavior of the polarized
PDFs. The lowest moments of the latter are known from $\beta$ decays
(see, for instance, Ref.\ \ci{BB}). The usual assumption of a smooth
behavior of the PDFs without a change of sign, leads to opposite signs
of $\Delta u_{\rm val}$ and $\Delta d_{\rm val}$. As our numerical
studies reveal the gluon and sea quark contributions to ${\cal M}^{U}$ 
are very small and compensate each other to a large extent since the
gluonic and sea quark GPDs $\widetilde{H}$ have opposite signs. Their 
combined contributions are practically negligible. This is the reason 
why we quote only the expansion parameters of $\tilde{h}_i$ for the 
valence quarks in Tab.\  \ref{tab:par-tilde}. 
 
\begin{table*}
\renewcommand{\arraystretch}{1.4} 
\begin{center}
\begin{tabular}{|c|| c |c | c | c|}
\hline
         & $\Delta u_{\rm val}$ & $\Delta d_{\rm val}$  
                        & $e^u_{\rm val}$ & $e^d_{\rm val}$ \\
\hline
$\tilde{\delta}$ & $0.48$ & $0.48$ & 0.48 & 0.48 \\ 
$\tilde{c}_0$    & $0.61+0.033\,L$ & $-0.320-0.040\,L$
         & $\phantom{-}2.204$& $-3.114$ \\
$\tilde{c}_1$    & $\phantom{-}0.410-0.377\,L$ & $-1.427-0.176\,L$ & 
                                $-2.204$ & $\phantom{-}8.096$ \\
$\tilde{c}_2$    & $5.10-1.21\,L$ & $\phantom{-}0.692-0.068\,L$& $0.0$
         & $-6.477$ \\
$\tilde{c}_3$    & $0.0$ & $0.0$ & $0.0$ & $\phantom{-}1.295$ \\
\hline
\end{tabular}
\end{center}
\caption{The parameters appearing in the expansion \req{pdf-exp} of 
the polarized PDFs and the forward limits of $E^a$. The latter are
taken from Ref.\ \ci{DFJK4}. The expansion \req{pdf-exp} provides a 
fit to the Bl\"umlein-B\"ottcher PDFs \ci{BB} in the range 
$10^{-2} \leq \beta \leq 0.5$ and $ 4\,\gev^2 \leq Q^2 \leq
40\,\gev^2$. The powers $\tilde{\delta}$ are kept fixed in the fits to the PDFs.}
\label{tab:par-tilde}
\renewcommand{\arraystretch}{1.0}   
\end{table*} 

\begin{figure}[t]
\begin{center}
\includegraphics[width=.44\tw,bb=8 307 535 753,%
clip=true]{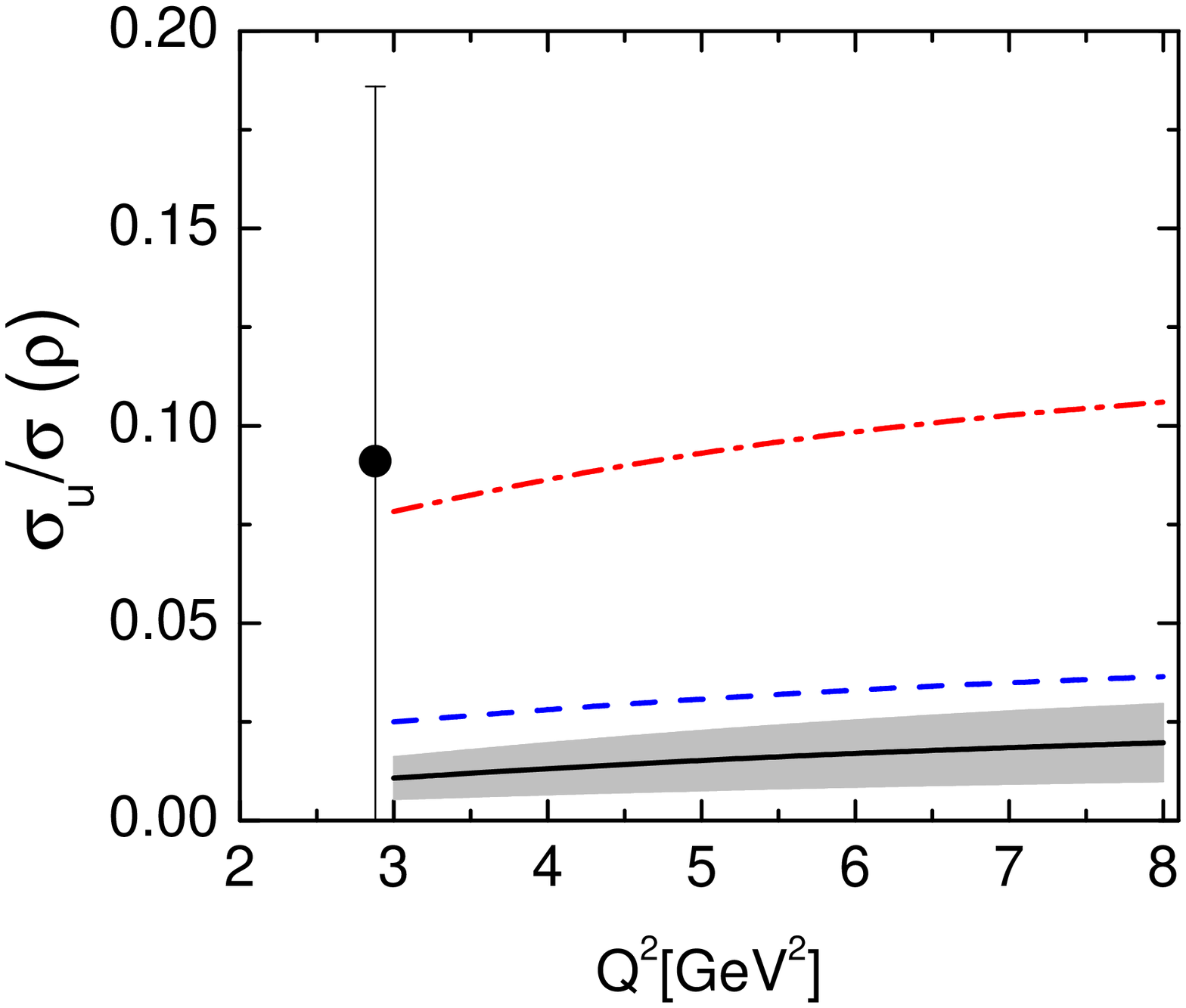} 
\includegraphics[width=.43\tw,bb=39 323 532 755,%
clip=true]{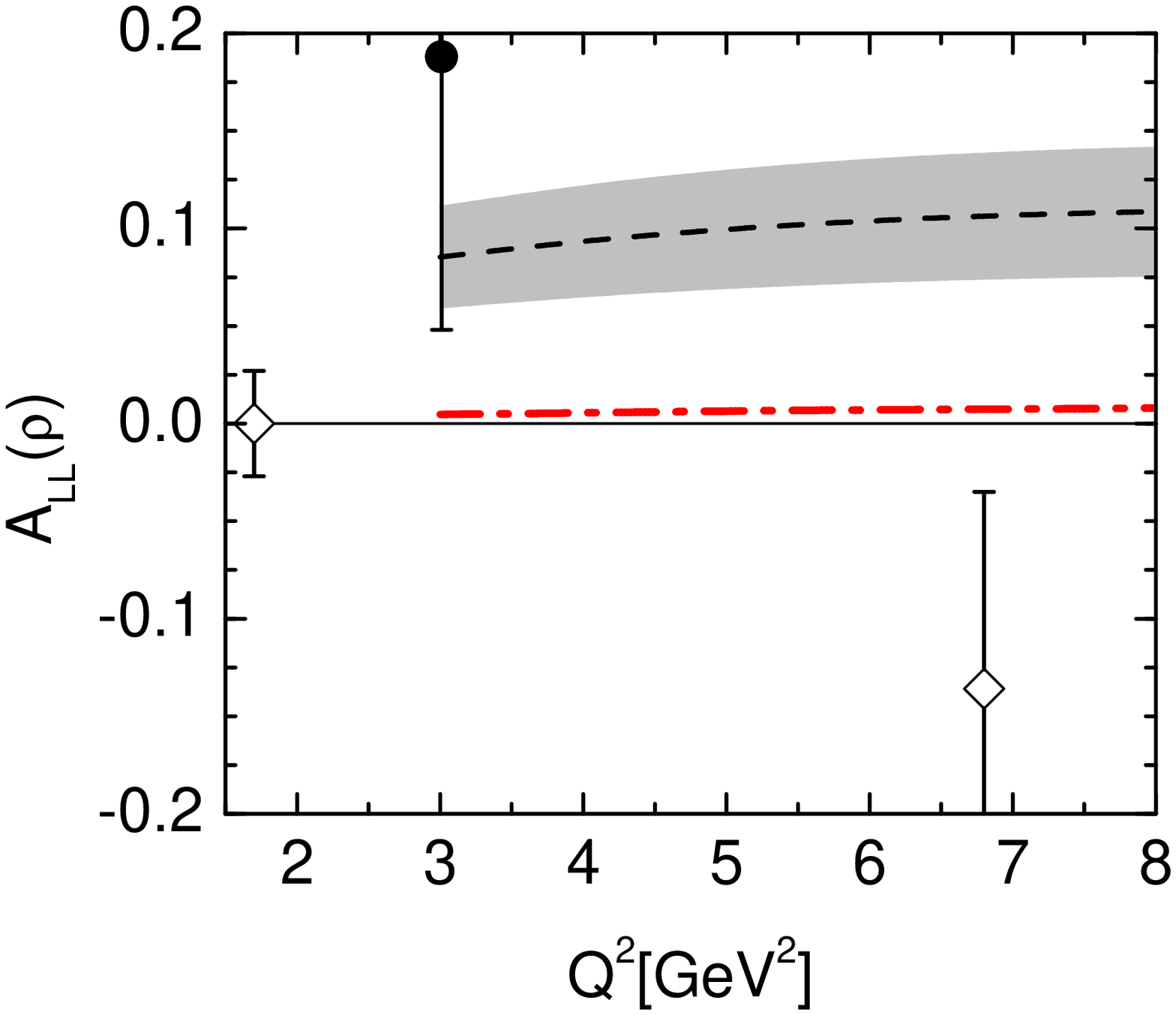} 
\end{center}
\caption{Left: The ratio of $\sigma_U$ and $\sigma$ for $\rho$ production
  versus $Q^2$ at $W=5\,\gev$. Data taken from HERMES \ci{hermes-draft}. 
  The solid (dashed, dash-dotted) line represents our estimate (obtained
  with $e_uH^u_{\rm val}+e_dH^d_{\rm val}$, with $e_uH^u_{\rm
    val}-e_dH^d_{\rm val}$), see text.
  Right: The helicity correlation $A_{LL}$ for $\rho$ production at $W=
  5\, (10)\,\gev$ dashed (dash-dotted) line. Data taken from COMPASS
  \ci{compass07} and HERMES \ci{hermes03}. For other notations cf. to
  Figs.\ \ref{fig:ratio1} and \ref{fig:ratio2}.}
\label{fig:sigmaU}
\end{figure}

Neglecting as in the preceding sections proton helicity flips and 
photon-meson transitions other than $L\to L$ and $T\to T$, one can 
project out the unnatural parity amplitude for $T\to T$ transitions
from a particular combination of SDMEs (see Eq.\ \req{eq:sdme}) 
\ba
U &=& \frac12\,\Big[1-r_{00}^{04} -2 r^1_{1-1}\Big] \nn\\
       &=&\frac{2}{N_T+\veps N_L} \, 
                   \mid {M}_{++,++}^{U}\mid^2\,.
\ea
This is the unnatural parity part of $N_T$ scaled by $N_T+\veps N_L$, see
\req{norm}. Integrating over $t$ one arrives at a cross section $\sigma_U$ 
defined in analogy to $\sigma_T$ in \req{sigma}. Evaluating this cross
section for $\rho$ production from the amplitudes given in \req{H-amp}
and \req{Ht-amp} and using the GPDs $\widetilde{H}$ described above,
we find the results shown in Fig.\ \ref{fig:sigmaU}. The cross section 
$\sigma_U(\rho)$ is rather small but in agreement with the preliminary 
HERMES result \ci{hermes-draft} at $Q^2=2.88\,\gev^2$ within an 
admittedly large error. For larger energies our approach will lead to
even smaller values for $\sigma_U$ since the valence quark contribution 
disappears and, as we mentioned above, the combined gluon and sea 
contribution is very small (the typical size of gluon plus sea quark 
contribution to $\sigma_U(\rho)$ is $0.013\,{\rm nb}$). 
We note that for $\rho$ 
production, the H1 data \ci{h1} provide values for $\sigma_U$ that 
are compatible with zero (e.g.\ at $Q^2=3\,\gev^2$, 
$\sigma_U/\sigma=0.03\pm 0.07$) while the ZEUS results \ci{zeus07} are 
about $1.5\sigma$ above zero (e.g.\ at $Q^2\simeq 3.4\,\gev^2$, 
$\sigma_U/\sigma=0.03\pm 0.02$). Both the experimental results are in 
agreement with our estimates within errors. An immediate consequence 
of the cancellation of gluon and sea contributions to the unnatural 
parity amplitude is that $\sigma_{U}$ for $\phi$ production is very 
small, in fact compatible with zero. This is in agreement with the 
preliminary HERMES data \ci{hermes-phi} and with the H1 data
\ci{adloff}. Thus, there is indication from both theory and experiment 
that  $\sigma_U$ for $\rho$ and $\phi$ production is small. Its
neglect in $\sigma_T$ seems to be justified. A larger cross
section $\sigma_U$ is to be expected for $\omega$ production
because the combination $e_u\widetilde{H}^u_{\rm val}+e_d
\widetilde{H}^d_{\rm val}$ occurs (see \req{flavor}) which is larger 
than $e_u\widetilde{H}^u_{\rm val} -e_d \widetilde{H}^d_{\rm val}$ 
given the relative sign of $\widetilde{H}^u_{\rm val}$ and  
$\widetilde{H}^d_{\rm val}$.

One may wonder whether or not it is possible to generate a value for
$\sigma_U$ as large as, say, the face value of the preliminary HERMES 
result \ci{hermes-draft} by using GPDs constructed from the polarized
PDFs via the double distribution ansatz. In order to examine this
issue we recall that the polarized PDFs are the difference of PDFs for 
helicity parallel and anti-parallel to that one of the proton while 
the unpolarized PDFs represent their sum. Suppose the gluon and sea 
quark contributions still cancel and assume now that the 
helicity-parallel distributions dominate which, in the limit $x\to 1$, 
follows from QCD \ci{brodsky94}. In this case the double distribution 
ansatz leads to $\widetilde{H}^a_{\rm val}=H^a_{\rm val}$. Admittedly 
this is an extreme example since in all analyses \ci{BB,LSS,AAC} the
polarized $d$-quark distribution is negative. Another, more moderate  
example is set by the assumption $\widetilde{H}^u_{\rm val}=H^u_{\rm
  val}$ and $\widetilde{H}^d_{\rm val}=-H^d_{\rm val}$, implying $d$ 
quarks with dominantly antiparallel helicity. The results obtained
from these two scenarios are also shown in Fig.\ \ref{fig:sigmaU}. 
Obviously, it seems difficult to obtain agreement with the HERMES 
result \ci{hermes-draft} with GPDs constructed from the double 
distribution ansatz except extreme scenarios are realized in nature. 
  
The size of the amplitude ${\cal M}^{U}$ can be elucidated further
by considering the initial state helicity correlation $A_{LL}$
which can be measured with a longitudinally polarized beam and target. 
In contrast to cross sections and SDMEs where the corrections are
bilinear in the ${\cal M}^U$ terms and, hence, very small, the
leading term in $A_{LL}$ is an interference between the ${\cal M}^N$ and 
the ${\cal M}^{U}$ amplitudes. In fact, with the help of
parity conservation as well as \req{sym-N} and \req{sym-U}, one 
obtains (see \ci{first}) 
\be
 A_{LL}[ep\to eVp]\= 4 \sqrt{1-\veps^2}\,
                \frac{{\rm Re}\;\Big[ {\cal M}^N_{++,++}\, 
                 {\cal M}^{U*}_{++,++}\Big]}
               {N_T + \veps N_L}\,. 
\label{eq:all}
\ee
We stress that here the target polarization is specified relative to
the virtual photon direction while in experiment it is usually defined
with respect to the lepton beam direction. The conversion from our
specification to the one used in experiments leads to a factor
$\cos{\theta_\gamma}$ in \req{eq:all} \ci{sapeta05} where
$\theta_\gamma$ denotes the angle describing the rotation in the
lepton plane from the direction of the incoming lepton to the one of
the virtual photon. This angle is given by \ci{sapeta05}
\be
\cos{\theta_\gamma} \= \sqrt{1-\gamma^2
  \,\frac{1-y-y^2\gamma^2/4}{1+\gamma^2}} \simeq 1 - \frac12\gamma^2
(1-y)\,.
\label{eq:conversion}
\ee
The two parameters appearing in \req{eq:conversion} are 
$y=(W^2+Q^2-m^2)/(s-m^2)$, one of the conventional
variables of electroproduction, and $\gamma=2\xbj m/Q$. In the
kinematical situation of interest $\gamma$ is very small and, hence,
$\cos{\theta_\gamma}\simeq 1$. According to \req{eq:TT-supp}  $A_{LL}$
is of order $\langle k^2_\perp\rangle/Q^2$ and, therefore, expected to
be very small. In Fig.\ \ref{fig:sigmaU} our results for $\rho$ 
production are shown at $W=5$ and $10\,\gev$. For the lower energy the 
valence quark contribution generates values for $A_{LL}$ of about 
0.1 while at $10\,\gev$ only extremely small values are 
found. The valence quark contribution has nearly disappeared at that 
energy and, as we mentioned above, the gluon and sea quark
contributions cancel each other to a large extent. For instance, at 
$W=5\,\gev$ and $Q^2=3\,\gev^2$, $A_{LL}(\rho)$ changes by $-0.002$ 
if the gluon and sea quarks are neglected. Because of the
very small combined gluon and sea contributions we also predict very 
small values of $A_{LL}(\phi)$. For instance, at 
$W=5 (10)\,\gev$ and $Q^2=3\,\gev^2$, $A_{LL}(\phi)=-0.002 (-0.007)$. 
 
Recently, the COMPASS collaboration has measured $A_{LL}$ for $\rho$
production \ci{compass07}. For $Q^2$ less than $2\,\gev^2$ COMPASS
observes very small values for $A_{LL}$, which are compatible with 
zero. Their only data point for which its $Q^2$ is sufficiently large 
for application of the handbag approach, is inconclusive because of 
its extremely large error; it is at variance with our results by 
about $1\sigma$, see Fig.\ \ref{fig:sigmaU}. The HERMES results on 
this observable for $\rho$ and $\phi$ production \ci{hermes03} is 
in agreement with our predictions. The SMC experiment \ci{SMC} 
observes a double spin asymmetry for $\rho$ production at $W=15\,\gev$
that is compatible with our results within large errors. In passing 
we remark that $A_{LL}$ is sensitive to the relative phase 
$\delta_{NU}$ between the amplitudes ${\cal M}^N_{++,++}$ and 
${\cal M}^U_{++,++}$. Therefore, a large value of $\sigma_U$ is not 
necessarily in contradiction with a small value of $A_{LL}$ provided 
the phase is near $90^\circ$. Nevertheless, in our approach the phase 
is small. For instance, at $W=5\,\gev$ and $Q^2=3\,\gev^2$ we find 
$\delta_{NU}=3.7^{\,\circ}$, i.e.\ a large value of $\sigma_U$ would go 
along with a large value of $A_{LL}$ in our approach. For instance, a
scenario with $\widetilde{H}^a_{\rm val}=H^a_{\rm val}$ leads to
$A_{LL}(\rho)=0.14$ for this kinematical point of reference.

Let us return to the issue of the size of the combined gluon and sea
contribution to the unnatural parity amplitude. Up to know we assumed
that this combined contribution is very small as follows from the
double distribution ansatz using current polarized PDFs. One may
wonder whether these PDFs are really correct or whether the smallness
of the combined gluon and sea contribution is perhaps a
special feature of our double distribution ansatz for $\widetilde{H}$.
First we note that in all current analyses of the polarized PDFs
\ci{BB,LSS,AAC} $\Delta g$ and the polarized sea have opposite signs
and are rather small in magnitude~\footnote{
A negative $\Delta g$ is also discussed in Ref.\ \ci{LSS}.}. 
In particular a large positive
$\Delta g$ is in conflict with measurements of the $A_{LL}$ asymmetry
in the production of jets \ci{star} and $\pi^0$ mesons \ci{phenix} in 
inclusive proton-proton collisions. A negative polarized sea is for
instance seen in the HERMES semi-inclusive deep-inelastic scattering  
data \ci{sidis}. Thus, we think that the main features of the polarized 
PDFs are correct. A small combined gluon and sea contribution from
$\widetilde{H}$ seems to be required by the relevant data too. Leaving
aside the HERMES results on $\sigma_U$ and $A_{LL}$ for $\rho$
production which receive contributions from $\widetilde{H}_{\rm val}$,
we note that all other pertinent data are small and in most cases compatible
with zero. These data are $\sigma_U(\rho)$ from H1 \ci{h1} and ZEUS
\ci{zeus07}, the same cross section for $\phi$ production from HERMES 
\ci{hermes-phi} and H1 \ci{adloff} and finally the $A_{LL}$ data from
Compass \ci{compass07} and SMC \ci{SMC}. Thus, scenarios in which 
the combined gluon and sea quark contribution is large in magnitude 
seem to be excluded by the current data. Our double distribution ansatz for
$\widetilde{H}^g$ and $\widetilde{H}_{\rm sea}$ , as naive it may be,
qualitatively reproduces the main features of the data.

\section{ Proton helicity flip}
\label{sec:hel-flip}
The analysis of observables for vector-meson electroproduction 
measured with a transversely polarized proton target requires 
the proton helicity flip amplitude. The handbag contribution
to this amplitude is given by
\be
{\cal M}_{\mu -, \mu +}^{Ni}(V) \=
             - \frac{e}{2}\,\frac{\sqrt{-t}}{2m}\, \sum_a 
         e_a{\cal C}_V^{\,a}\, \int_{\bar{x}_i}^1\, d\xb \, 
    \Big[{\cal H}^{Vi}_{\mu +,\mu +}+{\cal H}^{Vi}_{\mu -,\mu -}\Big] 
             \,E^i(\xb,\xi,t)\,,
\label{E-amp}
\ee
where $t_{\rm min}$ is ignored. Our choice of the phase of this
amplitude is in accord with conventions exploited in 
\ci{diehl07,sapeta05}. In general there is also a contribution from 
the GPD $\widetilde{E}$ feeding the amplitude ${\cal M}^{Ui}_{+-,++}$. 
It is expected to be small and neglected in our estimate of the size 
of the proton helicity flip amplitude. The evaluation of ${\cal M}^N$ 
for proton helicity flip is analogous to that of the non-flip amplitude
\req{H-amp} except that the convolution is now to be performed with
the GPD $E$ instead of $H$. The construction of $E$ through double 
distributions is also analogous to that of $H$, see Sect.\
\ref{sec:GPD}. The only but crucial difference is that the forward
limit $e(x)=E(x,0,0)$ is inaccessible in deep inelastic lepton-nucleon 
scattering. However, the forward limits of the valence quark GPDs have
been determined phenomenologically in the form factor analysis 
performed in Ref.\ \ci{DFJK4}. The parameters of $e^u_{\rm val}$ and 
$e^d_{\rm val}$ expanded according to \req{pol-pdf-exp}, are taken 
from \ci{DFJK4}. They are quoted in Tab.\ \ref{tab:par-tilde} at a 
scale of $4\,\gev^2$. For other scales these functions are unknown
which does not matter here since we will estimate proton helicity flip
only for photon virtualities near that value. Note that in contrast to 
$u_{\rm val}$ and $d_{\rm val}$, $e^u_{\rm val}$ and $e^d_{\rm val}$ 
have opposite signs. This is due to the fact that they are normalized by
\be
\int_0^1 dx\, e^a_{\rm val}(x) \= \kappa_a\,,
\ee
where $\kappa^a$ gives the contribution of quark flavor $a$ to the
anomalous magnetic moment of the proton ($\kappa_u \simeq 1.67$,
$\kappa_d \simeq -2.03$). The forward limits of $E$ for gluons and 
sea quarks are unknown. But there is a sum rule 
\be
\int_0^1 dx\,x\, e^g(x) \= -\sum_a \int_0^1 dx 
  x\, e^a_{\rm val}(x) -2\sum_a \int_0^1 dx x\, e^{\bar{a}}(x)\,,
\label{sum-rule}
\ee
which follows from a combination of Ji's sum rule and the momentum sum
rule of deep ineleastic lepton-nucleon scattering \ci{diehl03}.
Neglecting a possible difference between $e^s$ and $e^{\bar{s}}$, we
can evaluate the valence-quark term in the sum rule \req{sum-rule}
from the GPDs specified in Tab.\ \ref{tab:par-tilde}. We obtain
\be
\sum_a \int_0^1 dx x\, e^a_{\rm val}(x) \= 0.008\pm 0.007\,.
\label{moment2}
\ee
This signals a remarkable compensation between the second moments of 
$e^u_{\rm val}$ and $e^d_{\rm val}$ 
\be
\frac{\int_{0}^1 dx\, x\,[e^u_{\rm val}+e^d_{\rm val}]}
{\int_{0}^1 dx\, x\,[e^u_{\rm val}-e^d_{\rm val}]}  \simeq 0.026\,,
\ee
which is even stronger than that of their first moments
($(\kappa_u+\kappa_d)/(\kappa_u-\kappa_d)\simeq 0.1$). The error in 
\req{moment2} has been estimated from those quoted in \ci{DFJK4}. 
Hence, the moment of $e^g$ in \req{moment2} is only about as large as 
the sum of the sea quark moments. This is to be contrasted with the 
situation for $H$ where the corresponding gluon moment is more than
four times larger than the sum of the sea quark ones. Another argument 
that points into the same direction is the behavior of the gluon 
(or Pomeron) Regge trajectory. As is well-known this trajectory 
couples mainly to the proton helicity non-flip vertex while the flip 
coupling is very small. It is hard to find phenomenological evidence 
for a non-zero flip coupling \ci{donnachie}. Thus, the relative 
importance of gluon and valence quark GPDs is very different for $E$ 
and $H$. It seems unlikely that $E^g$ plays an analogously prominent 
role as $H^g$. In a first step we therefore assume that, for HERMES 
kinematics, proton helicity flip is dominated by the valence quarks 
(see also the discussion in Ref.\ \ci{kugler07}). 
Since the same Regge poles contribute to $E$ and $H$, we therefore 
use the standard valence-quark trajectory here as well and assume 
$b^e_{\rm val}=0$, too. With regard to this situation we cannot
estimate the size of proton helicity flip for $\phi$ production but we
expect it to be very small. We stress that due to the opposite signs
of $E^u_{\rm val}$ and $E^d_{\rm val}$, their contribution to $\rho$ 
production off protons, $\propto e_u E^u_{\rm val} - e_d E^d_{\rm val}$, 
is much smaller than that from the corresponding contribution of 
$H^a_{\rm val}$. This provides additional justification for the
neglect of $E$ in the proton helicity non-flip amplitude (see 
discussion after Eq.\ \req{xi-xbj}). An interesting case is $\omega$ 
production since for it the combinations $e_u E^u_{\rm val} + e_d
E^d_{\rm val}$ and $e_u H^u_{\rm val} + e_d H^d_{\rm val}$ occur. The 
first combination is larger, the second one smaller than for $\rho$ 
production and, hence, a markedly larger ratio of proton helicity flip 
and non-flip is expected for $\omega$ production. For instance, at 
$Q^2=4\,\gev^2$ and $W=5\,\gev$ the flip/non-flip ratio of the
absolute values of the $\omega$ amplitudes is about 13 times larger 
than the corresponding ratio for $\rho$ production.

\begin{figure}[t]
\begin{center}
\includegraphics[width=.43\tw,bb=8 319 534 752,%
clip=true]{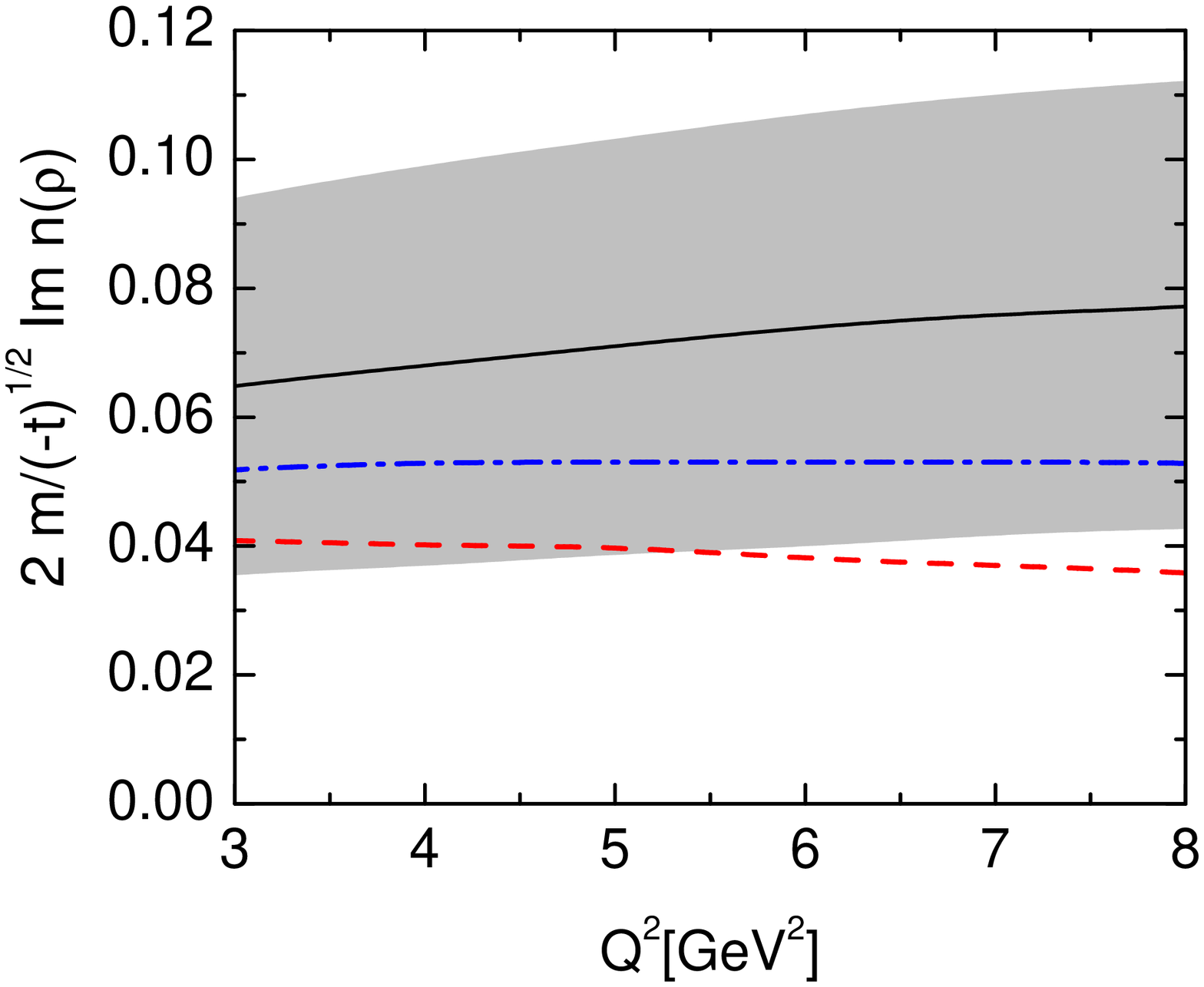} \hspace*{0.03\tw}
\includegraphics[width=.43\tw,bb=10 316 535 754,%
clip=true]{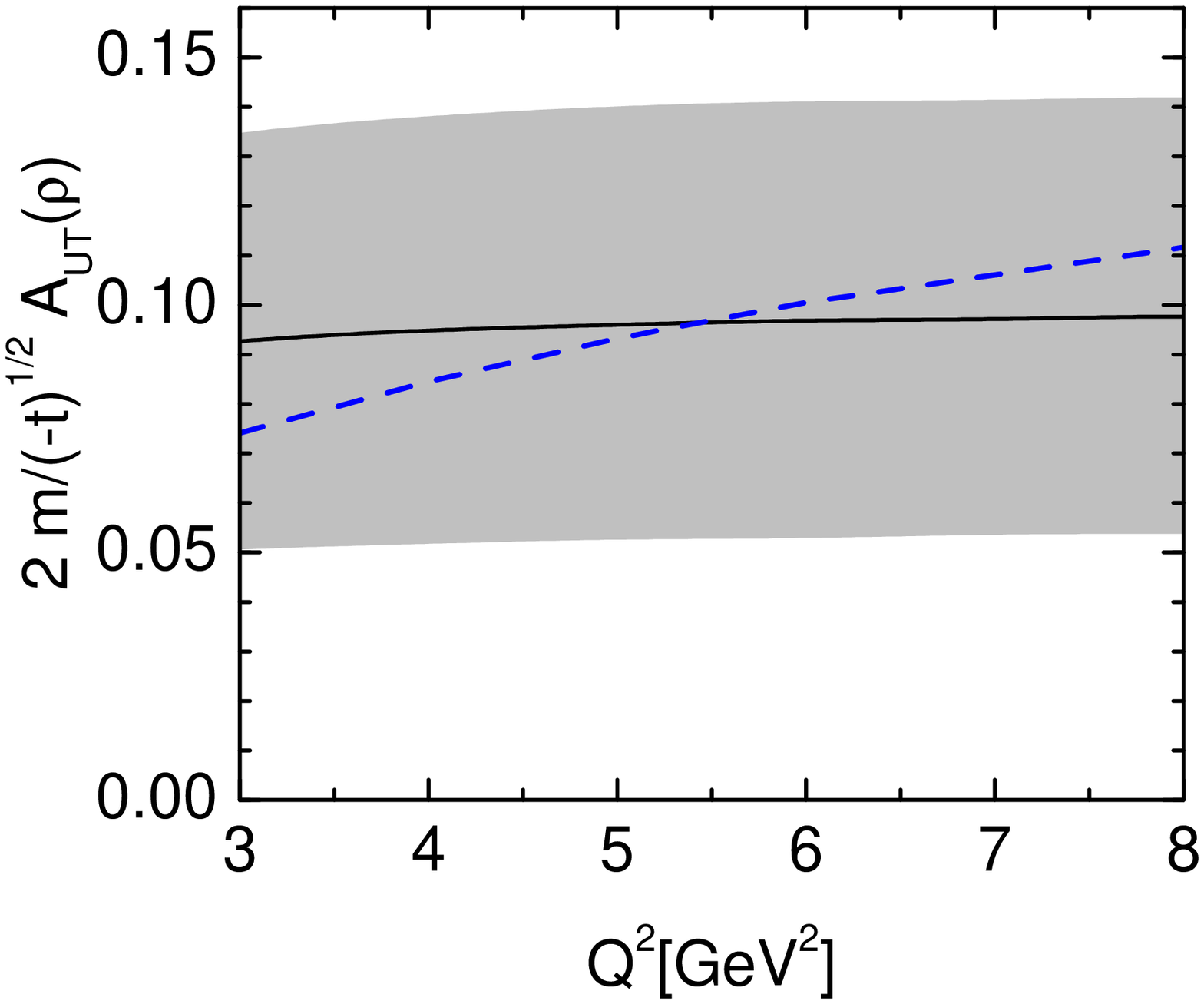} 
\end{center}
\caption{Left: The imaginary parts of the SDMEs $n_{00}^{00}$  (solid), 
$n_{++}^{++}$ (dashed) and $n^{0+}_{0+}$ (dash-dotted line), scaled by 
$2m/\sqrt{-t}$, for $\rho$ production versus $Q^2$ at $W=5\,\gev$. The 
error band is only shown for $n_{00}^{00}$. 
Right: The asymmetry $A_{UT}$,  scaled by $2m/\sqrt{-t}$, for $\rho$ 
production versus $Q^2$ at $W=5\,\gev$. The dashed line represents the
  leading-twist contribution.} 
\label{fig:sigmaN}
\end{figure}

Recently the formalism for the SDMEs in the case of a proton target 
polarized perpendicular ('normal') with respect to the plane in which 
the scattering $\gamma^* p\to Vp$ takes place, has been developed 
\ci{diehl07}. These SDMEs are denoted by $n^{\sigma\sigma'}_{\mu\mu'}$ 
and related to bilinear combinations of the amplitudes for helicities 
$\mu, \mu'$ and $\sigma, \sigma'$ of the virtual photon and the meson, 
respectively:
\be
n^{\sigma\sigma'}_{\mu\mu'} \= \frac1{N_T + \veps N_L} \sum_{\nu'} \left[
                  {\cal M}_{\sigma\nu',\mu +}\,{\cal M}^{*}_{\sigma'\nu', \mu' -}
 -{\cal M}_{\sigma\nu',\mu -}\, {\cal M}^{*}_{\sigma'\nu',\mu'
		    +}\right]\,.
\label{def-n}
\ee
If one neglects the amplitudes ${\cal M}^U$ in accord with our
findings described in Sect.\ \ref{sec:Htilde}, as well as helicity
flip transitions $\gamma^*\to V$ only the SDMEs 
\be
n^{\mu\mu'}_{\mu\mu'}\=\frac{2}{N_T + \veps N_L} \,\sum_{\nu'}
           {\cal M}^N_{\mu\nu',\mu+}\,{\cal M}^{N*}_{\mu'\nu',\mu'-}\,
\label{n-non-flip}
\ee
are non-zero. Explicitely these SDME read
\ba
n^{00}_{00} &=& \frac{4\,{\rm i}}{N_T+\veps N_L}\, {\rm Im}\Big[{\cal
      M}^N_{0-,0+}\,{\cal M}^{N*}_{0+,0+}\Big] \nn\\[0.2em]
n^{++}_{++}&=& n^{-+}_{-+} \=\frac{4\,{\rm i}}{N_T+\veps N_L}\,{\rm Im}\Big[{\cal
      M}^N_{+-,++}\,{\cal M}^{N*}_{++,++}\Big] \nn\\[0.2em]
n^{0+}_{0+} &=&- (n^{+0}_{+0})^*\=  \frac2{N_T+\veps N_L}\,  
    \Big[{\cal M}^N_{0-,0+}\,{\cal M}^{N*}_{++,++} - 
        {\cal M}^N_{0+,0+}\,{\cal M}^{N*}_{+-,++}   \Big]\,.
\label{eq:n}
\ea
For non-zero SDMEs $n_{\mu\mu'}^{\mu\mu'}$ phase differences between
the proton helicity flip and non-flip amplitudes are mandatory.
Such phase differences are provided by the handbag approach since the
non-flip amplitudes are built up by gluons and quarks while the
flip amplitudes receive only contributions from the valence quarks
in our model for the  GPD $E$. Indeed we obtain the values $38.8^\circ$ 
and $34.7^\circ$ for the phase between the proton flip and non-flip
amplitudes in the case of longitudinal and transverse photon
polarization, respectively. In Fig.\ \ref{fig:sigmaN} the SDMEs 
\req{eq:n} are shown versus $Q^2$ at $W=5\,\gev$, the trivial factor
$\sqrt{-t}/2m$, see \req{E-amp}, is pulled out and $t$ set to zero
otherwise. The scaled SDMEs are small, of the order of five percent. 
These SDMEs will be measured by HERMES.

One may also consider transverse proton polarization lying in the
$\gamma^*p\to Vp$ plane ('sideways'). In this case SDMEs, denoted by
$s^{\sigma\sigma'}_{\mu\mu'}$ \ci{diehl07}, occur that are analogous to
\req{def-n} but with a plus sign between the two terms. The SDMEs
$s^{\mu\mu'}_{\mu\mu'}$ for photon-meson helicity non-flip  
are given by products of two small amplitudes, ${\cal M}^N$ for proton
helicity flip and ${\cal M}^U$. They are therefore very small in our
approach.

Finally, we estimate the asymmetry $A_{UT}$ of $ep\to e Vp$ for a 
transversely polarized target, normal to the $\gamma^*p\to Vp$ 
scattering plane. It is measured as the $\sin{(\phi-\phi_S)}$ moment 
of the electroproduction cross section where $\phi$ is the azimuthal 
angle between the lepton and hadron plane and $\phi_S$ the azimuthal 
angle of the target spin vector defined with respect to the direction 
of the virtual photon \ci{diehl07}. As for the asymmetry $A_{LL}$ the
conversion of this spin vector into the one used in the experimental
setup where the target polarization is specified relative to the
lepton beam, again leads to a factor $\cos{\theta_\gamma}$ in
principle. According to the discussion following Eq.\
\req{eq:conversion} it is replaced by one. 
In the handbag approach the dominant contribution to this asymmetry reads
\be
A_{UT}(ep\to eVp) \= 4\; \frac{{\rm Im}\big[{\cal M}^N_{+-,++}\,{\cal
      M}^{N*}_{++,++}\big]
       +\veps\, {\rm Im}\big[{\cal M}^N_{0-,0+}\,{\cal
         M}^{N*}_{0+,0+}\big]}
         {N_T + \veps N_L}
\label{eq:AUT}
\ee 
It is just the imaginary part of the sum of $n_{++}^{++}$ and 
$\veps n_{00}^{00}$ and is also proportional to  $\sqrt{-t}/2m$. We
again pull out the latter factor and display the scaled asymmetry, 
evaluated at $t=0$, in Fig.\ \ref{fig:sigmaN}. We obtain a positive 
asymmetry. In contrast to $A_{LL}$ it is finite to leading-twist order
which is obtained from \req{eq:AUT} by neglecting the contributions from
transverse photons and evaluating those from longitudinal photons in
collinear approximation. For comparison the leading-twist contribution 
is also shown in Fig. \ref{fig:sigmaN}. It is not too different from
the full result. A preliminary HERMES result \ci{hermes-aut} for
$\rho$ production, integrated on the range $0\geq -t \geq 0.4\,\gev^2$, 
is $-0.033\pm 0.058$ at $Q^2=3.07\,\gev^2$ and $W=5\,\gev$ while we
find $0.02\pm 0.01$ for this kinematical situation. Note that the
scaled asymmetry is still $t$ dependent although mildly. Ignoring this
and integrating just $\sqrt{-t}$ one makes an error of about $10\%$ at
$Q^2\simeq 3-4\,\gev^2$. For $\omega$ production $A_{UT}$ is about 10 
times larger than for $\rho$ production. For $\phi$ production, on the 
other hand, we expect a very small asymmetry since the gluon and sea 
contributions are not only small but cancel each other to some extent,
see \req{sum-rule}.
  
The asymmetry $A_{UL}$ for an unpolarized beam and a longitudinally
polarized target is given by the same expression as $A_{UT}$. Only the
mentioned conversion factor $\cos{\theta_\gamma}$ is to be replaced
by $\sin{\theta_\gamma}$ which is very small \ci{sapeta05}. The beam asymmetry
$A_{LU}$ obtained with a logitudinally polarized beam and an
unpolarized target is zero given that helicity flip $\gamma^*\to V$
transitions can be neglected.

\section{Summary}
\label{sec:summary}
Together with Ref.\ \ci{second} this work gives an exhaustive
description of light vector-meson electroproduction within the
handbag approach for a wide range of kinematics reaching from the
HERMES up to the HERA kinematical settings. Our handbag approach  
includes power corrections which suppress the leading-twist amplitude 
for $\gamma^*_L\to V_L$ transitions and allows for a calculation of the
transverse amplitude $\gamma^*_T\to V_T$. In order to specify our
handbag approach fully we have to mention the soft physics input,
namely the GPDs that are constructed from PDFs with the help of double
distributions, and the light-cone \wf s for the mesons. GPDs and \wf s
affect the handbag amplitudes differently and can therefore be 
disentangled. The \wf s provide effects of order $\langle \vk^2 \rangle/Q^2$
controlled by the transverse size parameter $a_V$ while the GPDs
mainly influence the $\xbj$ dependence of the amplitudes or, at fixed
$Q^2$, the $W$ dependence. Besides the dominant
contributions from the GPD $H$ ('natural parity' contribution) we also
estimated effects from the GPDs $\widetilde{H}$ ('unnatural parity'
contribution) and $E$ controlling the proton helicity flip
amplitudes. These effects are generally small. With our analysis we
achieve a very good description of the HERA, HERMES and COMPASS data
on the separated and unseparated cross sections for $\rho$ and $\phi$
electroproduction, on the ratio $\sigma_L/\sigma_T$, on the SDMEs and
on some spin asymmetries. The only problem we observed is that the
relative phase between the longitudinal and transverse amplitudes
seems to be larger in experiment, in particular in the HERMES
experiment \ci{hermes-draft}, than our handbag approach predicts. The
neglect of contributions from transitions other than $\gamma^*_L\to
V_L$ and $\gamma^*_T\to V_T$ to the cross sections seems to be
justified. Only little contributions from $\gamma^*_T\to V_L$
transitions are to be observed in some of the SDMEs experimentally. 
We note that in
Ref.\ \ci{VGG} the longitudinal amplitude has also been analyzed
within the handbag approach. The main difference to our work is that
in \ci{VGG} the gluonic contribution is treated in leading-log
approximation \ci{koepf} and added incoherently to the quark
amplitudes. This line of action understimates the gluonic contribution
at low energies. 

The applicability of our approach is limited to $Q^2\gsim 3-4\,\gev^2$,
$W\gsim 4- 5\,\gev$ and $\xbj\lsim 0.2$. The restriction of $Q^2$ is
due to the mentioned - and still unsettled - difficulties with higher-order
perturbative corrections as well as due to the neglected 
corrections of order $m^2/Q^2$ and  $-t/Q^2$. There may be also power
corrections of other dynamical origin which become large at low $Q^2$.  
The restriction of $W$ has its origin in the asymmetric minimum the
cross sections exhibit at $W\simeq 3-4\,\gev$. While the cross sections
\ci{zeus07,h1,zeus98,e665} mildly increase towards larger $W$, they
\ci{hermes-draft,cornell,clas} increase sharply in the opposite
direction. In fact they rise by nearly an order of magnitude between
$W\simeq 4$ and $2\,\gev$. Obviously, a new dynamical mechanism seems
to set in and the handbag physics is perhaps not applicable here. On
the other side, it dominates for $W\gsim 4\,\gev$. The mild increase
of the cross sections with energy beyond the minimum is well described 
by the handbag physics as the results presented in this article and 
in Ref.\ \ci{second} reveal. The restriction of $\xbj$ is of
difference quality. It allows to neglect contributions of order
$\xbj^2$ or $\xi^2$, e.g.\ in \req{eq:H-E}, which simplifies the
analysis of vector-meson electroproduction strongly.

\section*{Acknowledgements} 
We thank A.\ Borissov, M.\ Diehl, N.\ d'Hose, A.\ Levy, W.-D.\ Nowak 
and A.\ Sandacz for discussions. We are also grateful to the
HERMES and COMPASS collaborations for permission to use preliminary
data. This work has been supported in part by the 
Russian Foundation for Basic Research, Grant 06-02-16215, the 
Integrated Infrastructure Initiative ``Hadron Physics'' of the 
European Union, contract No. 506078 and by the Heisenberg-Landau 
program. 

\end{document}